\documentclass[12pt]{article}

\usepackage{latexsym,amsmath,amssymb,epsfig,graphicx}

\newcommand{\pkt}{\; .}
\newcommand{\kma}{\; ,}
\newcommand{\nn}{\nonumber}
\newcommand{\bea}{\begin{eqnarray}}
\newcommand{\eea}{\end{eqnarray}}
\newcommand{\be}{\begin{equation}}
\newcommand{\ee}{\end{equation}}
\newcommand{\beast}{\begin{eqnarray*}}
\newcommand{\eeast}{\end{eqnarray*}}
\newcommand{\eqn}[1]{(\ref{#1})}

\newcommand{\calf}{{\cal F}}

\newcommand{\calm}{{\cal M}}
\newcommand{\caln}{{\cal N}}

\newcommand{\re}{{\rm Re}}
\begin{document}

\begin{titlepage}
\begin{flushright}
DO-TH-05/07 \\
hep-th/yymmnn \\
May 2005\\
revised: February 2006
\end{flushright}

\vspace{20mm}
\begin{center}
{\Large \bf
Tunneling in a quantum field theory on a compact one-dimensional space}
\vspace{10mm}

{\large
J\"urgen Baacke \footnote{e-mail: baacke@physik.uni-dortmund.de}
and
Nina Kevlishvili  
\footnote{e-mail: nina.kevlishvili@het.physik.uni-dortmund.de}}

\vspace{15mm}

{\large  Institut f\"ur Physik, Universit\"at Dortmund \\
D - 44221 Dortmund, Germany
}
\vspace{15mm}

{\bf Abstract}
\end{center}

We compute tunneling in a quantum field theory in 
$1+1$ dimensions for a field potential $U(\Phi)$ of
the asymmetric double well type. The system is localized initially 
in the ``false vacuum''. We consider the case of a {\em compact space}
($S_1$) and study {\em global} tunneling.  
The process is studied in real-time simulations.
The computation is based on the time-dependent Hartree-Fock
variational principle with a product {\em ansatz} for the
wave functions of the various normal modes. While the wave functions 
of the nonzero momentum modes are  treated within 
the Gaussian approximation, 
the wave function of the zero mode that tunnels between the two wells
is not restricted to be Gaussian, but evolves according to a 
standard Schr\"odinger equation. We find that in general tunneling
occurs in a resonant way, the resonances being associated with
degeneracies of the approximate levels in the two separated wells.
If the nonzero momentum modes of the quantum field are excited only weakly,
the phenomena resemble those of quantum mechanics with the wave function 
of the zero mode oscillating forth and back between the wells. 
If the nonzero momentum modes are excited efficiently, they react back onto
the zero mode causing an effective dissipation. In some region of
parameter space this back-reaction causes the
potential barrier to disappear temporarily or definitely, the tunneling 
towards the ``true vacuum'' is then replaced by a sliding of the wave function.

\end{titlepage}

\setcounter{page}{2}
\section{Introduction}
\label{intro}
\setcounter{equation}{0}
Tunneling is one of the important elementary processes
that may happen in a quantum mechanical system. There is a vast
literature on the subject and the WKB approach is discussed in
textbooks on quantum mechanics. In quantum field theory
tunneling has been mostly discussed as a local process proceeding
via bounce  or bubble solutions 
\cite{Coleman:1977py,Coleman:1980aw,Linde:1980tt,Linde:1981zj} 
for the simple reason that in an
infinite space a global transition of a mean field through a barrier 
would have an infinite action. However, if space is finite such global 
transitions are possible. 

Tunneling with compact spaces can be of interest in cosmology.
Such transitions in de Sitter space have been evoked in order 
to describe the quantum creation of a universe 
\cite{Hawking:1981fz,Hawking:1998bn}. In a less specific way one may 
think of transitions within a finite volume of a chaotic
initial state of the universe \cite{Linde:1983gd}.
Tunneling in a finite volume may also 
have some relation to local tunneling in an infinite
volume as an alternative to bounce transitions. 
This point has been elaborated in Ref. \cite{Hirota:2000kk}.

Tunneling with compact space also occurs in models
for the quantum creation of a universe \cite{Hartle:1983ai,Vachaspati:1988as}, 
which in the minisuperspace space approximation is a closed 
$3$ sphere and whose radius $a$ is a simple quantum mechanical
degree of freedom. This effect has been widely discussed,
and most authors use the WKB approximation
\cite{Vilenkin:1994rn,Vilenkin:1999pi,Hong:2002yf,Coule:1999wg,Bouhmadi-Lopez:2002qz,Kim:2004jj}.
Our analysis is not directly related to quantum cosmology because
the r\^ole of time is quite different there and here.
However, there are similarities concerning the  r\^ole of particle
production and their back-reaction which still are an open issue 
there\cite{Rubakov:1984pa,Levkov:2002qg,Hong:2002yf,Hong:2003un}.
 
Tunneling is of course a process that occurs in real time and
the continuation to imaginary time is a technical tool which
is widely used and whose application is based 
in a controlled way on the eikonal expansion. Its application
to quantum field theory becomes cumbersome whenever one is interested
in the physical evolution of the system during and after
the tunneling process, the matching of the wave functions
that is well understood in quantum mechanics now encompasses
an infinite system of modes and becomes rather involved.
It therefore seems to be of interest to look
for a description of the system entirely in real time.

Such an approach was taken for the case of {\em quantum mechanics}
in Ref. \cite{Nieto:1985ws}.
These authors have studied in detail the case of the
double-well potential. They find that tunneling is
characterized by the occurrence of resonances
between  degenerate (approximate) levels in the spectrum of
the separate wells.

For quantum field theory the  real time approach to quantum tunneling, 
in the same way as it is used here, has recently been discussed 
by Hirota \cite{Hirota:2000kk}. His analysis, as ours, is based
on the time-dependent Hartree-Fock approximation.
He uses analytic approximations for the zero mode wave functions.
While quantum back reaction and renormalization are mentioned,
they are not considered in detail and the author does not
present any numerical results.

We do not solve the field theoretical problem exactly but use 
two approximations: for the wave function of the entire system
we make a product {\em ansatz}, i.e., a Hartree approximation. 
The wave functions of the fluctuation modes are taken to be Gaussian; 
for the zero mode wave function, however, we numerically solve the 
Schr\"odinger equation. In this respect we differ from the widely 
used out-of-equilibrium simulations 
\cite
{Eboli:1987fm,Cooper:1997ii,Boyanovsky:1996sq,Baacke:1996se,Ramsey:1997sa}, 
where a Gaussian wave function, shifted by a classical
field $\phi(t)$, is used for the zero mode as well. 
By using the Hartree approximation we go beyond the one-loop back-reaction
by  taking into account the back-reaction
of the nonzero momentum modes onto the zero 
mode {\em and onto themselves}.
The latter back-reaction avoids a possible instability of the system
whenever the effective mass of the nonzero momentum modes
gets imaginary.
 
The plan of the paper is as follows: in Sec. \ref{model} we present 
the model, the decomposition into a discrete set of degrees
of freedom, and the Hamiltonian in the Schr\"odinger
representation; in Sec. \ref{hartree} we formulate the time-dependent
Hartree approximation \cite{Jackiw:1979wf,Cooper:1986wv,Destri:1999he} 
using our {\em ansatz} for the product wave function; renormalization 
is discussed in Sec.
\ref{renorm} and the initial conditions in
Sec. \ref{init}; numerical results for some parameter sets are presented
and discussed in Sec. \ref{numerics}; we end with conclusions and an outlook in
Sec. \ref{outlook}.  

 
\section{The model}
\setcounter{equation}{0}
\label{model}
We consider a scalar quantum field theory in $1+1$ dimensions with a compact space,
i.e., on a manifold $R\times S_1$. As ultimately we want to use our approach on
$R\times S_3$ we only introduce those self-couplings that would lead to a renormalizable
theory in $3+1$ dimensions. The action is given by
\be
S=\int_0^{2\pi a} dx \int dt\left[\frac{1}{2}\dot{\Phi}^2
-\frac{1}{2}\left(\frac{\partial\Phi}{\partial x}\right)^2
-U(\Phi)\right]
\kma\ee
with
\be
U(\Phi)=\frac{1}{2}m^2\Phi^2-\eta \Phi^3+\frac{\lambda}{8}\Phi^4
\ee
and periodic boundary conditions in $x$. With positive values of the
couplings the potential has the asymmetric double well form
with a minimum at $\Phi=\Phi_-=0$ and a second one at 
\be
\Phi_+=\frac{3\eta}{\lambda}\left(1+\sqrt{1-\frac{8}{9}\alpha}\right)
\kma\ee
with $\alpha=\lambda m^2/4\eta^2$. The parameter $\alpha$
is restricted to the range $0<\alpha< 1$.
The field $\Phi$ is dimensionless, and
$\eta$ and $\lambda$ have the dimension $mass^2$.
We introduce the expansion into normal modes
\be
\label{normalmodes}
\Phi(x,t)=\varphi_0(t)+\sum_{n=-\infty}^\infty\varphi_n(t)e^{ik_n x}
\kma\ee
with 
\be
k_n=\frac{n}{a}
\pkt\ee
Then the action takes the form
\bea
S&=&2\pi a \int dt \left[\frac{1}{2}\dot\varphi_0^2+
\frac{1}{2}\sum_{n\neq0}\dot\varphi_n\dot\varphi_{-n}
-\frac{1}{2}\sum_n \omega_n^2\varphi_n\varphi_{-n}\right.
\\\nn
&& \left.+\eta \sum_{nn'}\varphi_n\varphi_{n'}\varphi_{-n-n'}
-\frac{\lambda}{8}\sum_{nn'n''}\varphi_n\varphi_{n'}\varphi_{n''}\varphi_{-n-n'-n''}\right]
\kma\eea
with
\be
\omega_n^2=m^2+\frac{n^2}{a^2}
\pkt \ee
As the field $\Phi$ is real we have the condition $\varphi_{-n}=\varphi_n^\dagger$.
We therefore introduce, for $n\neq 0$  real fields via
\be
\varphi_{\pm n}=\frac{1}{\sqrt 2}(\varphi_{n1}\pm i \varphi_{n2}),\; \;\; n>0
\pkt\ee 
In the Schr\"odinger representation the
 Hamiltonian is given by
\bea
H&=&2\pi a\left\{-\frac{1}{8\pi^2a^2}\left[\frac{\partial^2}{\partial\varphi_0^2}
+\sum_{n>0,j}\frac{\partial^2}{\partial\varphi_{nj}^2}\right]\right.
\\\nn
&&+\frac{1}{2}\left[m^2\varphi_0^2+\sum_{n> 0,j}\omega_n^2\varphi_{nj}^2\right]
\\\nn
&&\left.-\eta \sum_{nn'}\varphi_n\varphi_{n'}\varphi_{-n-n'}
+\frac{\lambda}{8}\sum_{nn'n''}\varphi_n\varphi_{n'}\varphi_{n''}
\varphi_{-n-n'-n''}\right\}
\pkt\eea
In order to get rid of factors $2\pi a$ we introduce the rescaling
\be \label{canonicalscale}
\varphi_k=\chi_k/\sqrt{2\pi a}
\kma\ee
so that the Hamiltonian takes the form
\bea \label{canonicalscalehamiltonian}
H&=&-\frac{1}{2}\left[\frac{\partial^2}{\partial\chi_0^2}
+\sum_{n>0,j}\frac{\partial^2}{\partial\chi_{nj}^2}\right]
\\\nn
&&+\frac{1}{2}\left[m^2\chi_0^2+\sum_{n> 0,j}\omega_n^2\chi_{nj}^2\right]
\\\nn
&&-\eta' \sum_{nn'}\chi_n\chi_{n'}\chi_{-n-n'}
+\frac{\lambda'}{8}\sum_{nn'n''}\chi_n\chi_{n'}\chi_{n''}
\chi_{-n-n'-n''}
\kma\eea
with $\eta'=\eta/\sqrt{2\pi a}$ and $\lambda'=\lambda/2\pi a$.
Note that in the cubic and quartic parts we still have retained the 
complex fields 
\be
\chi_{\pm n}=\frac{1}{\sqrt 2}(\chi_{n1}\pm i \chi_{n2})
\pkt\ee


\section{The time dependent Hartree-Fock approximation}
\setcounter{equation}{0}
\label{hartree}
We now make a variational  {\em ansatz} for the wave function and impose a 
variational
principle, which is known under the name 
of ``time-dependent Hartree-Fock approach''.
The {\em ansatz} for the wave function is
\be
\Psi(\chi_0,\chi_n,t)=\psi_0(\chi_0,t)\prod_{n>0}\psi_n(\chi_{nj},t)
\pkt\ee
Furthermore we will restrict the {\em ansatz} for the modes with $n\neq0$ to a
Gaussian wave function
\be\label{gaussianansatz}
\psi_n(\chi_{nj},t)=\frac{e^{-i\alpha_n(t)}}{[2\pi\sigma_n^2(t)]^{1/4}}
\exp\left[-\frac{1}{2}\left(\frac{1}{2\sigma_n^2(t)}-
i\frac{s_n(t)}{\sigma_n(t)}\right)
\chi_{nj}^2\right]
\kma\ee
while we do not further specify $\psi_0$.
The time dependent variational principle
\cite{Dirac:1930,Jackiw:1979wf}, now imposes the condition
\be
\delta\int d\chi_0\prod_{n>0,j}d\chi_{nj}\Psi^\dagger(\chi_0,\chi_{nj},t)
\left(i\partial_t-H\right)\Psi(\chi_0,\chi_{nj},t)=0
\pkt\ee
We will find later that the dynamics is independent of  $j$,
which therefore has already been suppressed in the index for the wave 
functions.
Before we write down the resulting Schr\"odinger equations for the
various degrees of freedom we rewrite the Hamiltonian in an appropriate way.

The part of $H$ which exclusively contains the zero mode is given by
\be \label{H0canonical}
H_{00}=-\frac{1}{2}\frac{\partial^2}{\partial\chi_0^2}+\tilde U(\chi_0)
\kma\ee
 with
\be \label{potentialcanonical}
\tilde U(\chi_0)=\frac{1}{2}m^2\chi_0^2-\eta' \chi_0^3
+\frac{\lambda'}{8}\chi_0^4
\pkt\ee

The parts bilinear in the quantum modes lead to the Hamiltonian
\be
H_{0n}=\sum_{n>0,j}\left[-\frac{1}{2}\frac{\partial^2}
{\partial \chi_{nj}^2}+\frac{1}{2}\omega_n^2\chi_{nj}^2\right]
\pkt\ee

Due to the Gaussian {\em ansatz} for the $n\neq0$ modes the expectation values of 
any odd powers of $n\neq 0$ fluctuations vanish. So if we consider the
interaction between zero and nonzero modes we need to retain only terms 
bilinear in the $n\neq 0$  modes. We then have for the interaction
between zero and nonzero modes
\be
\left[\sum_{n,n'}\chi_n\chi_{n'}\chi_{-n-n'}\right]_{I0n}= 
3\chi_0\sum_{n\neq 0}\chi_n\chi_{-n}=
3\chi_0\sum_{n>0,j}\chi^2_{nj}
\ee
and
\be
\left[\sum_{n,n',n''}\chi_n\chi_{n'}\chi_{n''}
\chi_{-n-n'-n''}\right]_{I0n}
=
6 \chi_0^2\sum_{n\neq0}\chi_n\chi_{-n}=
6 \chi_0^2\sum_{n>0,j}\chi_{nj}^2
\pkt\ee

So with the scaled couplings $\eta' $ and $\lambda'$ the interaction 
Hamiltonian 
coupling zero and nonzero modes is given, in the approximation considered here,
by
\be
H_{I0n}=(-3\eta'\chi_0+\frac{3\lambda'}{4}\chi_0^2)\sum_{n>0,j}\chi_{nj}^2
\pkt\ee
The part of the Hamiltonian cubic in the nonzero modes has a vanishing 
expectation value.
The quartic term has the expectation value
\be
<\sum_{n,n',n''}\chi_n\chi_{n'}\chi_{n''}\chi_{-n-n'-n''}>
=3 \sum_{nj,n'j'}<\chi_{nj}^2><\chi_{n'j'}^2>
\kma\ee
where we have used the fact that the wave functions for $n\neq 0$
are Gaussian.
This part yields the Hamiltonian for the mutual and self interaction
of the nonzero modes. It is given by
\be
<H_{Inn}>=\frac{3}{8}\lambda'\left[\sum_{nj}<\chi_{nj}^2>\right]^2
= \frac{3}{2}\lambda' \calf^2
\pkt\ee
 Applying now the variation principle we get for $\psi_0$ the Schr\"odinger
equation 
\bea \label{schroe_zero}
i\partial_t \psi_0(\chi_0,t)=\left[
-\frac{1}{2}\frac{\partial^2}{\partial\chi_0^2}+\tilde U(\chi_0)+
(-6\eta'\chi_0+\frac{3\lambda'}{2}\chi_0^2)\calf\right]\psi_0(\chi_0,t)
\pkt\eea
Here we have introduced the fluctuation 
integral 
\be
\calf=\frac{1}{2}<\sum_{n>0,j}\chi_{nj}^2>
\kma\ee 
which will be specified more explicitly later.
For the modes with $n\neq0$ we find
\be
i\partial_t \psi_n(\chi_n,t)=\left[-\frac{1}{2}
\frac{\partial^2}{\partial \chi_{nj}^2}
+\frac{1}{2}(\omega_n^2+W(t))\chi_{nj}^2\right]\psi_n(\chi_n,t)
\pkt
\ee
Here  
\be\label{wdef}
W(t)=<\tilde U''(\chi_0)-m^2>+3 \lambda'\calf
\kma\ee
with
\be
<\tilde U''(\chi_0)-m^2>=\int d\chi_0\psi_0^*(\chi_0,t) 
[-6\eta'\chi_0+\frac{3\lambda'}{2}\chi_0^2]
\psi_0(\chi_0,t)
\pkt\ee
We also introduce $\Omega^2_n(t)=\omega^2_n+W(t)$.
The Schr\"odinger equation for the quantum mode wave function
is seen to be independent of the index $j$. 
Therefore the previous summations over
$j$ just lead to a degeneracy factor $2$, so that for the
mode functions the index $j$ 
can be suppressed.

With the Gaussian {\em ansatz} \eqn{gaussianansatz} the Schr\"odinger equation
for the $n\neq0$ modes 
implies
\bea 
\dot \sigma_n(t)&=&s_n(t)\kma
\\
\dot s_n(t)&=&-\Omega_n^2(t) \sigma_n(t)+\frac{1}{4\sigma^3_n(t)}\kma
\\
\dot\alpha_n(t)&=&\frac{1}{4\sigma^2_n(t)}\pkt
\eea
The first two of these equations can be related to mode functions
$f_n(t)$ satisfying 
\be
\ddot f_n(t) +\Omega_n^2(t)f_n(t)=0
\kma\ee
as they arise from a Klein-Gordon equation for a field with the effective
mass $m^2+W(t)$.
We have, with $\omega_{n0}=\Omega_n(0)$, 
\bea
\sigma_n(t)&=&\frac{|f_n(t)|}{\sqrt{2\omega_{n0}}}\kma
\\
s_n(t)&=&\frac{1}{\sqrt{2\omega_{n0}}}\frac{d}{dt}|f_n(t)|
\kma\eea
while the wave function is given by
\be\label{wavefn_modes}
\psi_n(\chi_n,t)=e^{-i\alpha_n(t)}
\left[\frac{2\omega_{n0}}{2\pi |f_n(t)|^2}\right]^{1/4}
\exp\left[\displaystyle\frac{i}{2}\frac{\dot f_n^*(t)}{f_n^*(t)}\chi_n^2\right]
\pkt
\ee
In deriving these relation one uses repeatedly the Wronskian relation
\be
\dot f_n^*(t) f_n(t)-f_n^*(t)\dot f_n(t) =2 i \omega_{n0} 
\kma\ee
which corresponds to an initial condition
\be
f_n(0)=1  \;\;,\;\; \dot f_n(0)=-i\omega_{n0}
\pkt\ee

For this wave function the expectation value of $\chi_n^2$ is given by
\be
\int d\chi_n\left|\psi_n(\chi_n,t)\right|^2\chi_n^2=
\frac{|f_n(t)|^2}{2\omega_{n0}}
\kma\ee
so that the fluctuation integral becomes
\be
\calf(t)=\sum_{n>0}\frac{|f_n(t)|^2}{2\omega_{n0}} 
\pkt
\ee

The system of equations we have presented here is consistent with
an conserved energy
\be
E=<H_{00}+H_{0n}+H_{I0n}+H_{Inn}>
\kma\ee
with
\bea\label{class_engy}
<H_{00}>&=&\int d\chi_0\psi_0^*(\chi_0,t) 
\left[-\frac{1}{2}\frac{\partial^2}{\partial \chi_0^2}+\tilde U(\chi_0)\right]
\psi_0(\chi_0,t)\kma
\\ \label{fluc_engy}
<H_{0n}>&=&2\sum_{n>0}\frac{1}{2\omega_{n0}}
\left[\frac{1}{2}\left|\dot f_n(t)\right|^2
+\frac{1}{2}\omega_n^2\left|f_n(t)\right|^2\right]\kma
\\\label{inter_engy}
<H_{I0n}>&=&<\tilde U''(\chi_0)-m^2>\calf\kma
\\\label{self_engy}
<H_{Inn}>&=&\frac{3}{2}\lambda'\calf^2\pkt
\eea
It is convenient to replace
$\omega_n^2$ by  $\omega_{n0}^2$ in $<H_{0n}>$, so that
$<H_{0n}>$ becomes the free Hamiltonian for the initial 
fluctuation wave functions.
Then the term $<H_{I0n}>$ receives an additional contribution $-W(0)\calf$.

A particle number for the $n\neq 0$ modes may be defined in various
ways. A Fock space is defined by the mode decomposition
\eqn{normalmodes}. To the operators $\chi_{nj}$ and their conjugate
momenta $\pi_{nj}=\partial/\partial \chi_{nj}$ we can associate
creation and annihilation operators referring to an oscillator
of frequency $\omega_{n0}$ via
\bea
c_{nj}&=&\sqrt{\frac{\omega_{n0}}{2}}\left(\chi_{nj}+\frac{1}{\omega_{n0}}
\frac{\partial}{\partial \chi_{nj}}\right)\kma
\\
c^\dagger_{nj}&=&\sqrt{\frac{\omega_{n0}}{2}}
\left(\chi_{nj}-\frac{1}{\omega_{n0}}
\frac{\partial}{\partial \chi_{nj}}\right)
\eea
and a particle number
\be
\caln_{nj}=<c^\dagger_{nj}c_{nj}>=
<\frac{1}{\omega_{n0}^2}\left[-\frac{1}{2}\frac{\partial^2}{\partial
\chi_{nj}^2}+\frac{1}{2}\omega_{n0}^2\chi_{nj}^2\right]>
\;;\ee
computing the expectation value with the wave function
$\psi_{n}(\chi_{nj})$, Eq. \eqn{wavefn_modes}, one finds
\be \label{particlenumbers}
\caln_{nj}=\frac{1}{4\omega_{n0}^2}\left[\left|\dot f_n(t)\right|^2
+\omega_{n0}^2\left|f_n(t)\right|^2\right]-\frac{1}{2}
\pkt\ee
In the present context the quantum excitations do not
describe free particles; we merely use the total ``particle number'' 
$\caln=\sum_{nj}\caln_{nj}$ as an indicator for the excitations of the
$n>0$ oscillators.


\section{Renormalization}
\setcounter{equation}{0}
\label{renorm}
The fluctuation integral and the energy density, as introduced 
in the previous section, are divergent quantities.
In $1+1$ dimensions the renormalization is in principle rather
straightforward. However, here it has to be done for a nonequilibrium system
with discrete momenta, and for the Hartree approximation, i.e., for a 
resummed perturbation series. We here limit ourselves to present
the main steps needed to derive the formulas that enter the numerical codes. 

Though we treat the system in a {\em nonperturbative}
approach, the divergences are related exactly to those
of standard perturbation theory.
The mode functions can be expanded perturbatively
with respect to the potential $V(t)$ which contains the couplings
$\eta $ and $\lambda$; so such an expansion is at the same
time an expansion with respect to these couplings.

As discussed in previous publications on
nonequilibrium dynamics \cite{Baacke:1996se}  we write the mode functions as
\be
f_n(t)=e^{-i\omega_{n0} t}\left[ 1 + h_n(t)\right]
\ee
and convert the differential equation for the $f_n(t)$
into an integral equation for the functions $h_n(t)$ as
\be
h_n(t)=\frac{i}{2\omega_{n0}}\int_0^t dt'
\left(e^{2i\omega_{n0} (t-t')}-1\right)V(t')\left[1+h_n(t')\right]
\pkt\ee
Here $V(t)=W(t)-W(0)$, see Eq. \eqn{wdef}.

We further have
\be
\left|f_n(t)\right|^2=1+2 \re h_n(t)+\left|h_n(t)\right|^2
\pkt\ee
One easily finds that $\re h_n(t)$ behaves as
\be
\re h_n(t)\simeq -\frac{V(t)}{4\omega_{n0}^2} 
+ O\left(\frac{1}{\omega_{n0}^3}\right)
\kma\ee
for large $n$. For the fluctuation integral one finds
\be
\calf=\sum_n\frac{1}{2\omega_{n0}}\left[1-\frac{V(t)}{2\omega_{n0}^2} 
+  O\left(\frac{1}{\omega_{n0}^3}\right)\right]\kma
\ee
so that we can separate it into a divergent and a subtracted finite
part as
\bea
\calf^{(0)}&=&\sum_{n=0}^\infty\frac{1}{2\omega_{n0}}\kma
\\
\calf_{\rm sub}&=&\sum_{n=0}^\infty\frac{1}{2\omega_{n0}}
\left[2\re h_n(t)+
\left|h_n(t)\right|^2\right]
\pkt\eea
The divergent part has to be regularized and to be separated
into the standard renormalization part and a finite contribution.
The sum over discrete momenta here and below can be done in the same way as
the Matsubara sums in finite temperature quantum field theory,
or using the Plana formula \cite{Hardy}.
One finds
\be
\sum_{n=1}^\infty\frac{1}{2\omega_{n0}}=\pi a\int_{-\infty}^\infty
\frac{dk}{(2\pi)2\omega_0}\left[1 + \frac{2}{\exp(2\pi a \omega_0)-1}
-\frac{1}{\pi a\omega_0}\right]
\kma \ee
with $\omega_0=\sqrt{k^2+m_0^2}$.
The first term in the bracket is obviously the divergent part we
would obtain for infinite space and goes into the renormalization of
the various couplings. The second term in the
bracket arises from the periodic boundary conditions. The third
part arises from the fact that $\calf$ contains the nonzero
modes only and no subtraction has been applied to the zero mode.
The first term can be regularized in a standard way \cite{Baacke:1996se};
 we have
\be
\left(\int\frac{dk}{(2\pi)2\omega_0}\right)_{\rm reg}=
\int\frac{d^{2-\epsilon}k}{(2\pi)^{2-\epsilon}}
\frac{i}{k_0^2-k^2-m_0^2+io}=\frac{1}{4\pi}L_{\epsilon 0}\kma
\ee 
with $L_{\epsilon 0}=2/\epsilon+\ln(4\pi/m_0^2)-\gamma$
using dimensional regularization. $m_0^2$ depends on the initial
conditions, and the renormalization condition should not; we therefore
write $L_{\epsilon 0}=L_\epsilon+\ln m^2/m_0^2$ with 
$L_\epsilon=2/\epsilon+\ln(4\pi/m^2)-\gamma$
and include the second term into the finite part. 
So we define the remaining finite part as
\be
\calf^{(0)}_{\rm fin}=2\pi a\int_{-\infty}^\infty\frac{dk}{(2\pi)2\omega_0}
\frac{1}{\exp(2\pi a \omega_0)-1} -\frac{1}{4m_0}+\frac{a}{4}
\ln\frac{ m^2}{m_0^2}
\ee
and the finite part of the fluctuation integral as
\be
\calf_{\rm fin}=\calf^{(0)}_{\rm fin}+\calf_{\rm sub}
\pkt\ee

We similarly decompose the fluctuation energy as 
\bea \label{energydecomp}
<H_{0n}>&=&2\sum_{n>0}\frac{1}{2\omega_{n0}}
\left[\frac{1}{2}\left|\dot f_n(t)\right|^2
+\frac{1}{2}\omega_n^2\left|f_n(t)\right|^2\right]
\\\nn
&=& E_{\rm fl, sub}-\frac{1}{2}m_0+a\int_0^\infty dk \omega_0
-2 a\int_0^\infty\frac{k^2 dk}{\displaystyle \omega_0\left(
\exp(2\pi a \omega_0)-1\right)}\kma
\eea
with
\be
 E_{\rm fl, sub}=2\sum_{n>0}
\frac{1}{2\omega_{n0}}
\left[\frac{1}{2}\left|\dot f_n(t)\right|^2
+\frac{1}{2}\omega_n^2\left|f_n(t)\right|^2-\omega_0^2\right]
\pkt\ee
Integrating by part the last (``thermal'') integral
in Eq. \eqn{energydecomp} can be recast into the form
of a free energy 
\be
-2 a\int_0^\infty\frac{k^2 dk}{\displaystyle \omega_0\left(
\exp(2\pi a \omega_0)-1\right)}=
\int_0^\infty\frac{dk}{2\pi}\ln\left(1-\exp(-2\pi a \omega_0)\right)
\pkt\ee
The integral over $\omega_0$ is given, in dimensional regularization, as
\be
\left[a \int_0^\infty dk\omega_0\right]_{\rm reg}=
\frac{a}{4}m_0^2\left(L_\epsilon + \ln \frac{m^2}{m_0^2}+1\right)
\ee
and we define
\be
E_{\rm fl,fin}=E_{\rm fl,sub}
+\frac{a}{4}m_0^2\left(\ln \frac{m^2}{m_0^2}+1\right)
+\int_0^\infty\frac{dk}{2\pi}\ln\left(1-\exp(-2\pi a \omega_0)\right)
\pkt\ee

The renormalization is done in analogy to the case of nonequilibrium dynamics 
in \cite{Baacke:2001zt}, following the scheme of Ref. \cite{Nemoto:1999qf},
 by adding a counterterm
\be
\delta C \calm^2=\delta C (m^2 +<\tilde U''(\chi_0)-m^2> +3\lambda' \calf_{\rm fin})
\pkt\ee
Here one has used already the finite gap equation 
\be \label{gap_t}
\calm^2=m^2 +<\tilde U''(\chi_0)-m^2> +3\lambda' \calf_{\rm fin}
\kma\ee
which determines the effective mass of the fluctuations.
Choosing 
\be
\delta C=-\frac{a}{4}L_\epsilon
\kma \ee
and defining the initial mass $m_0$ by the initial gap equation
\be\label{gap0}
m_0^2=m^2 +<\tilde U''(\chi_0)-m^2> +3\lambda' \calf_{\rm fin}^{(0)}
\kma \ee
the gap equation \eqn{gap_t} is satisfied for all $t$. It can be written as
\be
\calm^2(t)=m_0^2+V(t)
\kma\ee
with the previously introduced potential
\be
V(t)=<\tilde U''(\chi_0)> - m_0^2 +3 \lambda' \calf_{\rm fin}(t)
\kma\ee
which now is well-defined.

For the energy we obtain
\bea\nn
E&=&<H_{00}(\chi_0)> + E_{\rm fl,fin} -\frac{1}{2}m_0+ <\tilde U''-m_0^2>\calf_{\rm fin}
+\frac{3}{2}\lambda' \calf_{\rm sub}^2
\\&& +\frac{a}{4}m_0^2(\ln\frac{m^2}{m_0^2}
+1)+\int_0^\infty\frac{dk}{2\pi}\ln\left(1-\exp(-2\pi a \omega_0)\right)
\pkt\eea
The last two terms are, for a fixed radius $a$, independent of time; 
we leave them out when presenting the energy conservation.
The term $-m_0/2$ marks the absence of the zero mode in the
sum over fluctuations. If $m_0=m$ this term exactly cancels, at $t=0$,
the expectation value $<H_{00}(\chi_0)>$.
Likewise, the term $-1/4m_0$ in the initial gap equation \eqn{gap0} 
is cancelled,
for $m_0=m$, by the term $<\tilde U''(\chi_0)-m^2>$. Indeed if $a$ 
is sufficiently large, $am > 1$, corresponding to a ``temperature'' $T_{\rm eff}/m < 1/2\pi$,
the ``thermal integrals'' become very small; then the solution
$m_0$ of the gap equation indeed is close to $m$ and  $E\approx 0$.


\section{Initial conditions}
\setcounter{equation}{0}
\label{init}

As already discussed in Sec. \ref{hartree} the system is started 
for the nonzero modes with
\be
f_n(t)\simeq e^{-\omega_{n0} t}
\kma
\ee
which is equivalent to initial wave functions
\be
\psi(\chi_n,0)=\left[\frac{\omega_{n0}}{\pi}\right]^{1/4}e^{\displaystyle
-\omega_{n0}\chi_n^2/2}
\pkt\ee
Here $\omega_{n0}=\sqrt{m_0^2+n^2/a^2}$ and $m_0$ is determined by the
gap equation \eqn{gap0}.

 For the zero mode we likewise start with a Gaussian wave function
\be
\psi(\chi_0,0)=\left[\frac{m}{\pi}\right]^{1/4}e^{\displaystyle
-m\chi_n^2/2}
\pkt\ee
This would be the ground state wave function of an oscillator with
frequency $\omega=m$, i.e., the ground state wave function
for a potential with $\eta=\lambda=0$. 

The choice of these initial conditions is of course quite arbitrary.
Here we try to start in an  approximate ground state 
for the system ``in the left well'' of the double well potential.
We find the total energy of the system to be close to zero,
as one  would expect for such an approximate ground state (it 
is understood that the zero point energies of the nonzero modes 
are not included). A different choice would in general 
lead to a higher total energy and make tunneling easier, or make
the transition to a large part an ``over the barrier'' transition.


\section{Numerical results}
\setcounter{equation}{0}
\label{numerics}

\subsection{Numerical details}

We have implemented the formulas of the previous sections numerically.
This is essentially straightforward. The Schr\"odinger 
equation for the zero mode becomes a system of first
order differential equations for the values $\psi_0(\chi_{0,k})$ where
$\chi_{0,k}$ are the equidistant discrete values of the variable 
$\chi_0$, and the coupling arises from the discretized
second derivative 
\be
\frac{\partial^2}{\partial \chi_0^2}\psi_0(\chi_{0,k})
\to \frac{\psi(\chi_{0,k+1})+\psi(\chi_{0,k-1})-2\psi(\chi_{0,k})}
{(\Delta \chi_0)^2}
\kma\ee
where $\Delta \chi_0$ is the step width for the $\chi_{0,k}$.
The Schr\"odinger equation for the $n\neq 0$ modes is converted
into the second order mode equations, which are coupled to the
zero mode and the other nonzero modes by $V(t)$.
This discretization leads to instabilities in the time evolution
unless the time intervals are chosen of the order $\Delta \chi_0 ^2$.
We used $4000$ grid points for the zero mode $\chi_0$, which typically extends 
over a region of $-10 < \chi_0 < 50$. So $\Delta \chi_0$ is of the order
$10^{-2}$. The Runge-Kutta time step was chosen $\Delta t = .00002$.
This choice makes the codes very slow, much slower than those of
nonequilibrium dynamics in the Gaussian approximation. As we do not
use the Gaussian approximation for the zero mode wave function,
we have to compute, at each Runge-Kutta step, not only the 
sums over the quantum modes $\chi_n$, but also the averages
of various observables in the ground state wave function.
The excited quantum modes were taken into account up to
$n=200$. A simulation of the time evolution
until $t \simeq 100$ takes a few hours on a standard PC.   

The accuracy of the computations was checked by computing Wronskians
and energy conservation. The relative accuracy obtained was better than 
six significant digits. The energy conservation also checks
the correct implementation of the basic equations.
We show a typical example, for $\lambda=\eta=1$ (set $IV$ below)
and $a=1.2$. The total energy is $E=(2.39\pm 0.001)\times 10^{-5}$ throughout
the total time interval, the single components
take values up to $35$.

\begin{figure}
\begin{center}
\includegraphics[scale=0.4]{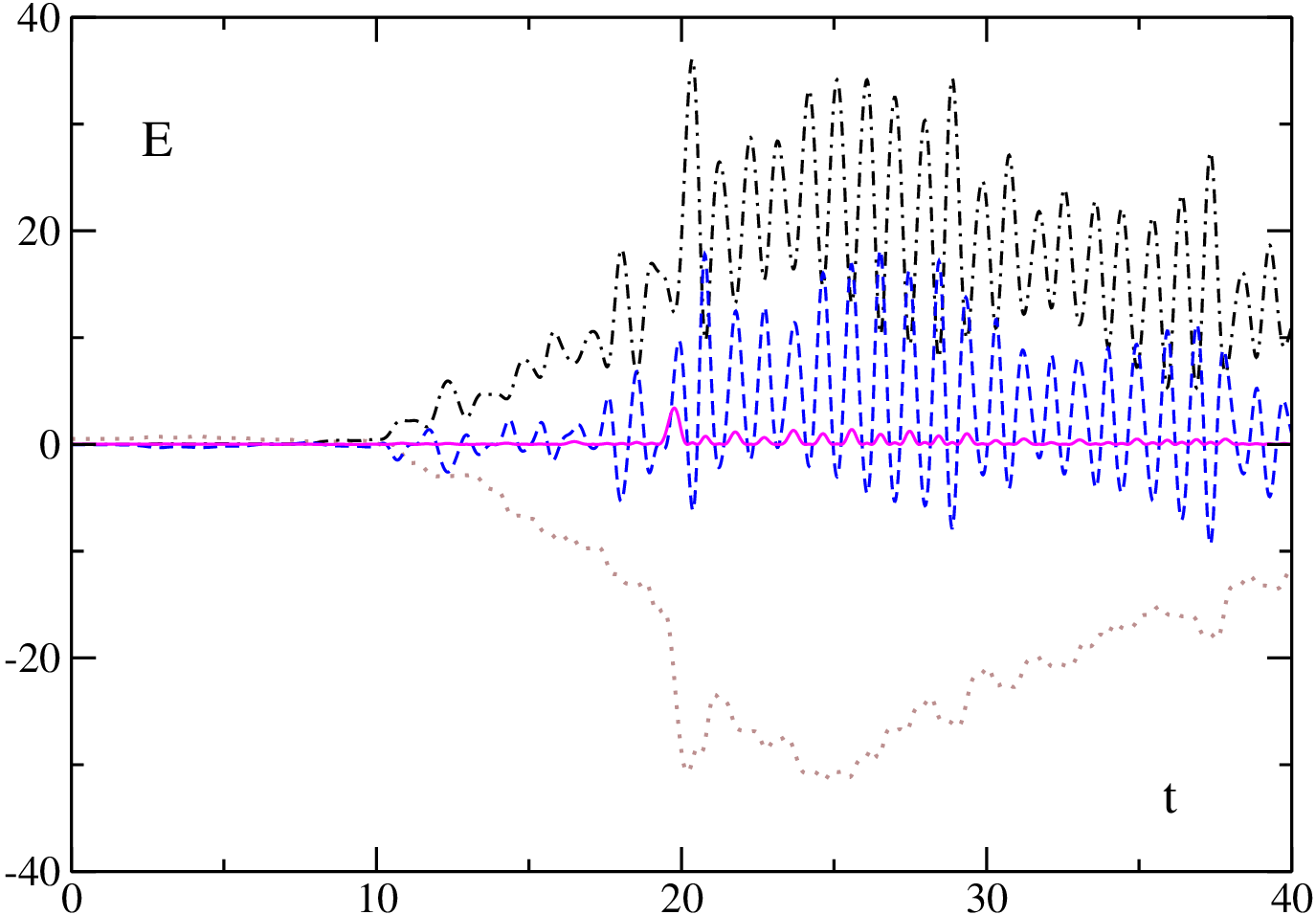}\vspace*{3mm}
\end{center}
\caption{\label{fig:energy}
Energy conservation; parameters $m=\eta=\lambda=1$, $a=6$;
 dotted line: classical energy, 
Eq.\eqn{class_engy};
dash-dotted line: fluctuation energy, Eq.\eqn{fluc_engy}; dashed line:
interaction energy, Eq.\eqn{inter_engy}; solid line:
self-interaction of the fluctuations, Eq.\eqn{self_engy}. The total
energy (not displayed) is $E=(2.39\pm 0.001)\times 10^{-5}$. }    
\end{figure}


\subsection{Parametrization and parameter sets}

The parameter space, encompassing $\eta,\lambda$ and $a$ is quite
large, so before presenting numerical results we try to 
get some qualitative insight into the physics to be expected
in certain ranges of their numerical values. The parameter
$m$ sets the overall scales and is put equal to $m=1$.

While in Sec. \ref{model}, we introduced a scaling of the fields,
\eqn{canonicalscale}, which was suitable for the canonical formalism
on the basis of the  Hamiltonian \eqn{canonicalscalehamiltonian}, the
discussion of the results is more transparent if we introduce the
$\alpha-\beta-$ parametrization widely used in  bounce computations
\cite{Linde:1981zj,Dine:1992wr,Baacke:1993ne,Baacke:2003uw}.
One introduces the rescaling $X = m x$ and $\Phi =m^2\hat \Phi/2\eta$.
Then for infinite space the classical action takes the form
\be
S_{\rm cl}=\frac{m^4}{4\eta^2}\int d^2 X \left[\frac{1}{2}
\left(\nabla_X \hat\Phi\right)^2 +\hat U(\hat\Phi)\right]
= \beta \hat S_{\rm cl}(\alpha)
\kma \ee
with $\beta = m^4/4\eta^2$, $\alpha = \lambda \beta/m^2$ and
\be
\hat U(\hat\Phi)=\frac{1}{2}\hat\Phi^2-\frac{1}{2}\hat\Phi^3
+\frac{\alpha}{8}\hat\Phi^4
\pkt\ee
While $\beta$ multiplies the classical action the one-loop 
effective action  is a function of $\alpha$ only.
So for large $\beta$ the system essentially becomes classical and the
quantum effects only lead to small corrections while for small $\beta$ 
the quantum effects become large. The parameter 
$\alpha$  determines the shape of the
potential: for $\alpha \to 1$ we get a symmetric double-well
potential, for small $\alpha$ the right hand well becomes much deeper 
than the left hand one, the barrier between the wells becomes shallow.
For bubble nucleation $\alpha \to 1$ is the thin-wall limit, where
the bubble size becomes very large.The case  $\alpha > 1$ is of no 
interest here, as  the minimum at $\Phi=0$ then becomes the global minimum.

We will in the following consider parameter sets with fixed $\alpha$ and
$\beta$, i.e., fixed $\lambda$ and $\eta$, and study the 
dependence on $a$.

For finite space extension the semiclassical  tunneling action of a
spatially homogeneous bounce is given by
\be \label{homogeneousbounceaction}
S_{\rm bounce}=2\pi a \beta \int_0^{\hat \Phi_0}d \hat \Phi
\sqrt{\hat U(\hat\Phi)} 
\kma \ee
where 
\be
\hat\Phi_0 =\frac{2}{\alpha}\left(1-\sqrt{1-\alpha}\right)
\ee
is the zero at the right hand side of the potential barrier.
For fixed $\alpha,\beta$ the tunneling via a homogeneous bounce 
should shut off with increasing $a$. The tunneling will then occur via
local bounces, as considered recently in the Hartree
approximation in \cite{Bergner:2003id,Baacke:2004xk}.

The transition rate obtained from the homogeneous bounce, Eq.
\eqn{homogeneousbounceaction}, 
will only have
a qualitative relation to the observed tunneling phenomena which will
be characterized by the occurrence of {\em resonances}. 
For quantum mechanical tunneling this was already observed in
Ref.\cite{Nieto:1985ws}. In order to estimate the separation
of the resonances we consider the approximate spectra of the 
left and right wells, treating them as separate oscillators.
The resonances can then be thought of as arising from the degeneracy
of levels in the left and in the right wells.
From the Hamiltonian and the potential written in canonical variables, 
Eqs. (\ref{H0canonical}) and (\ref{potentialcanonical}),
we see that the energies of the left hand oscillator are 
$E_n^l=(n+1/2)m$. The right hand oscillator potential has its minimum 
at 
\be
\chi_0^+=\sqrt{2\pi a}\Phi_+=\sqrt{2\pi a}\frac{2\eta}{\lambda}
\hat \Phi_+ 
\kma \ee
with 
\be
\hat \Phi_+=\frac{3}{2\alpha}\left(1+\sqrt{1-\frac{8}{9}\alpha}
\right)
\kma \ee
and the energy levels are given by
\be
E_n^r=\left(n+\frac{1}{2}\right)\omega_+ + \tilde U(\chi_0^+)
\pkt\ee
Here 
\be 
\omega_+^2=\tilde U''(\chi_0^+)=
2 m^2\left[\frac{9}{8\alpha}\left(1+\sqrt{1-\frac{8}{9}\alpha}
\right)-1\right]
\ee
and
\be
\tilde U(\chi_0^+)=2\pi a \beta \hat U(\hat\Phi_+)< 0
\pkt\ee  
So the condition for a degeneracy of a level in the spectrum of the right
hand well with the ground state of the left hand well
(our initial state) is
\be
2\pi a U(\Phi_+)+\left(n+\frac{1}{2}\right)\omega_+=\frac{1}{2}m
\pkt\ee
We present here the dependence of the tunneling phenomena on the
spatial length scale $a$. If the degeneracy holds for some value
$a$ and an integer $n$, it will hold again for 
 $a'=a+\Delta a$, $n'=n-\Delta n$ with
\be \label{resonancespacing}
\Delta a = - \frac{\omega_+}{2\pi U(\Phi_+)}\Delta n
\pkt\ee
The constant multiplying $\Delta n$
on the right hand side determines the spacing of resonance levels
as a function of $a$ at fixed $\alpha$ and $\beta$. 
Of course we may also have resonances between 
{\em excited}
states of the left well and those of the right hand well, but,
as our initial state will roughly correspond to the ground state
of the left hand well, these will be less important. Another, more essential
feature is the excitation of field quanta of the nonzero modes.
These will have a dissipative effect on the dynamics of the
zero mode and broaden the resonances. The interaction consists in a
deformation of the potential in which the zero mode is moving,
which takes the form
\be \label{effectivezeromodepotential}
U_0(\chi_0,t)=\tilde U(\chi_0)+(\tilde U''(\chi_0)-m^2)\calf(t)
\pkt\ee
If $\calf(t)$ is negligible the evolution of the system
proceeds like in the quantum mechanics of the zero mode.
If $\calf(t)$ remains small this time-dependent modulation of 
the potential will allow a few other approximate eigenstates 
of the zero mode to mix in,
the resonant oscillations develop ``higher harmonics'' and become 
irregular. If $\calf(t)$ gets large then the potential can be deformed
in such a way that the potential barrier disappears entirely, in such cases
the zero mode may ``slide'' into the new minimum. This happens if
$\calf(t)$ is positive; if $\calf(t)$ is negative the potential
is tilted counterclockwise and the barrier is enhanced. 


\subsection{Results of the numerical simulations}

We have performed a study of  tunneling as a function of the 
radius of the space manifold $S_1$, for fixed values
of the parameters $\alpha$ and $\beta$, or $\lambda$ and $\eta$;
the mass is chosen to be unity, which determines length and time scales.
We have considered four parameter sets:
set $I$: $\alpha=0.8,\beta=0.5$, i.e., $\lambda=1.6, \eta=1/\sqrt{2}$;
set $II$: $\alpha=0.6, \beta=2$, i.e., $\lambda=0.3, \eta=1/\sqrt{8}$;
set $III$: $\alpha=0.4, \beta= 1$, i.e., $\lambda=0.4, \eta = 1/2$;
set $IV$: $\alpha=0.25, \beta = 0.25$, i.e., $\lambda=\eta=1$.
For set $II$ we expect the quantum corrections to be small,
for sets $I$ and $IV$ we expect large quantum corrections, and 
moderate ones for set $III$.  

We have studied in general the behavior of the average
of the zero mode $<\chi_0 (t)>$, of the fluctuations (fluctuation
integrals, particle number, and the various energies), and of
the wave function $\psi_0(\chi_0)$ of the zero mode, as functions of time.
As the main indicator of the general behavior we use the 
maximal value attained by the expectation value of  the zero mode 
during the time evolution. We denote this value as
$\bar \varphi_0$. In those cases where we have effective tunneling, 
this zero mode average settles at a value of $\varphi_0$ beyond 
the potential barrier, the late time average being typically $20 \%$ 
smaller than the maximal value.

We display in Figs. 1 to 4 the dependence
of the maximal values  $\bar\varphi_0$ of $<\varphi_0(t)>$, as functions of
$a$ for the four parameter sets.
The horizontal lines labelled by $\Phi_m$ and $\Phi_0$ indicate
the position of the maximum of $U(\Phi)$ and the zero of the
position between the two minima (the ``end of the tunnel'').
 We see in all cases that an effective 
tunneling occurs as a resonance phenomenon. According to our estimate
in the previous subsection, Eq. \eqn{resonancespacing},  the spacing of 
different resonances as a function of $a$ would be given by
$\Delta a = 0.529   $ for set $I$, $\Delta a = 0.0338   $ for set $II$,
$\Delta a = 0.0167  $ for set $III$ and $\Delta a= 0.0166   $ for set $IV$.
The observed spacings roughly correspond to these estimates.
We note that also the excited states of the left hand oscillator
or of the nonzero modes can come into play, so if we observe some 
irregular spacings this may be due to such effects.
When compared to similar figures obtained in quantum mechanical
simulations \cite{Nieto:1985ws} the resonances seen in our simulations 
are broader. We observe that even on resonance the maximal value
of $<\varphi_0>$ never reaches the second minimum (the ``true vacuum'').
In part this is due to the fact that the effective potential seen by the
zero mode is modified by the quantum fluctuations, moreover, even
at late times the wave function generally retains a finite probability 
density near the false vacuum $\varphi_0=0$.
This will be discussed in more detail below.

\begin{figure}
\begin{center}
\includegraphics[scale=0.4]{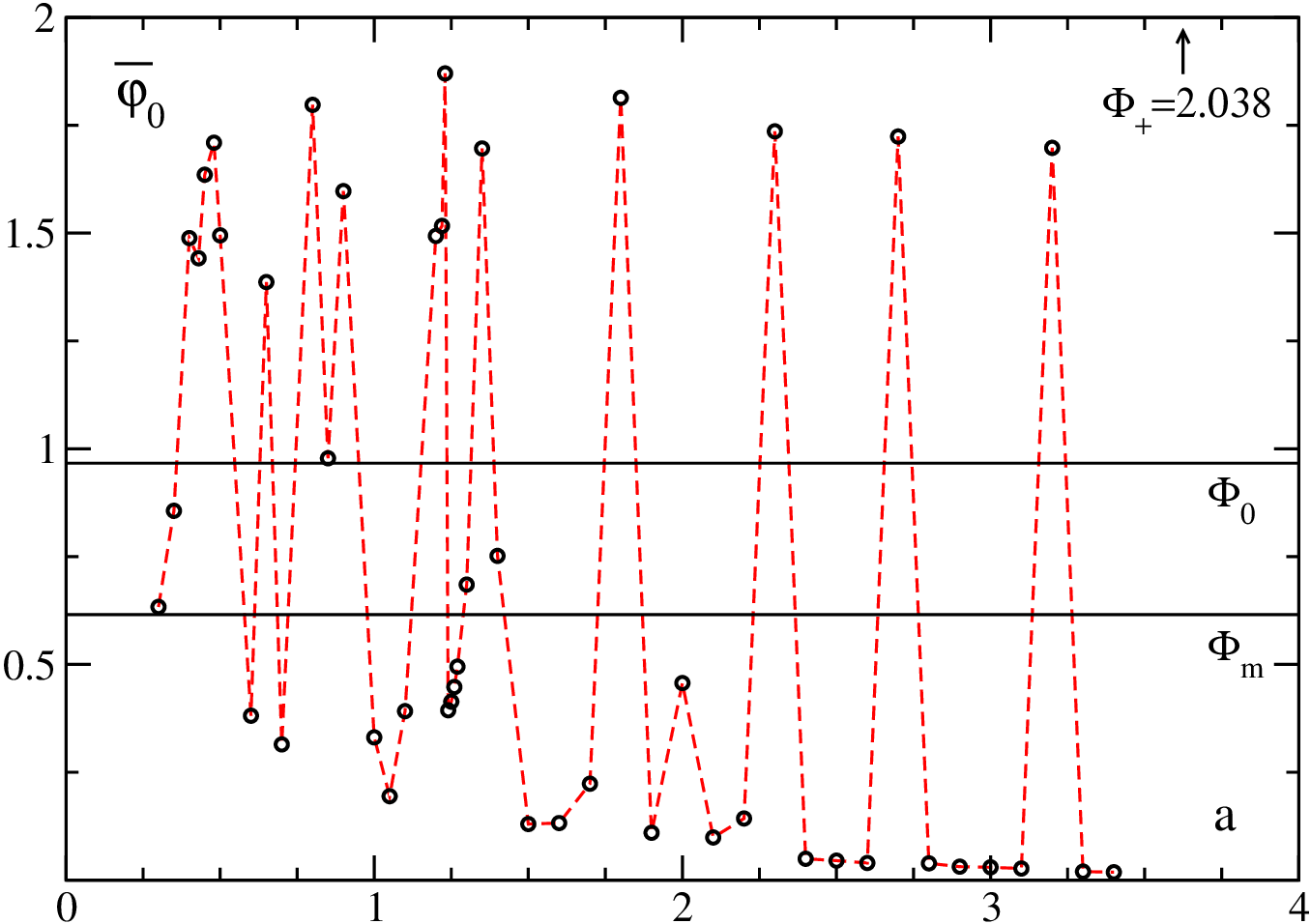}
\vspace*{3mm}
\end{center}
\caption{\label{fig:0.8series}
The maximum $\bar \varphi_0$ of the expectation value
 $<\varphi_0(t)>$ for set $I$,
$\alpha=0.8,\beta=0.5$, as a function of $a$.}
\end{figure}

\begin{figure}
\begin{center}
\includegraphics[scale=0.4]{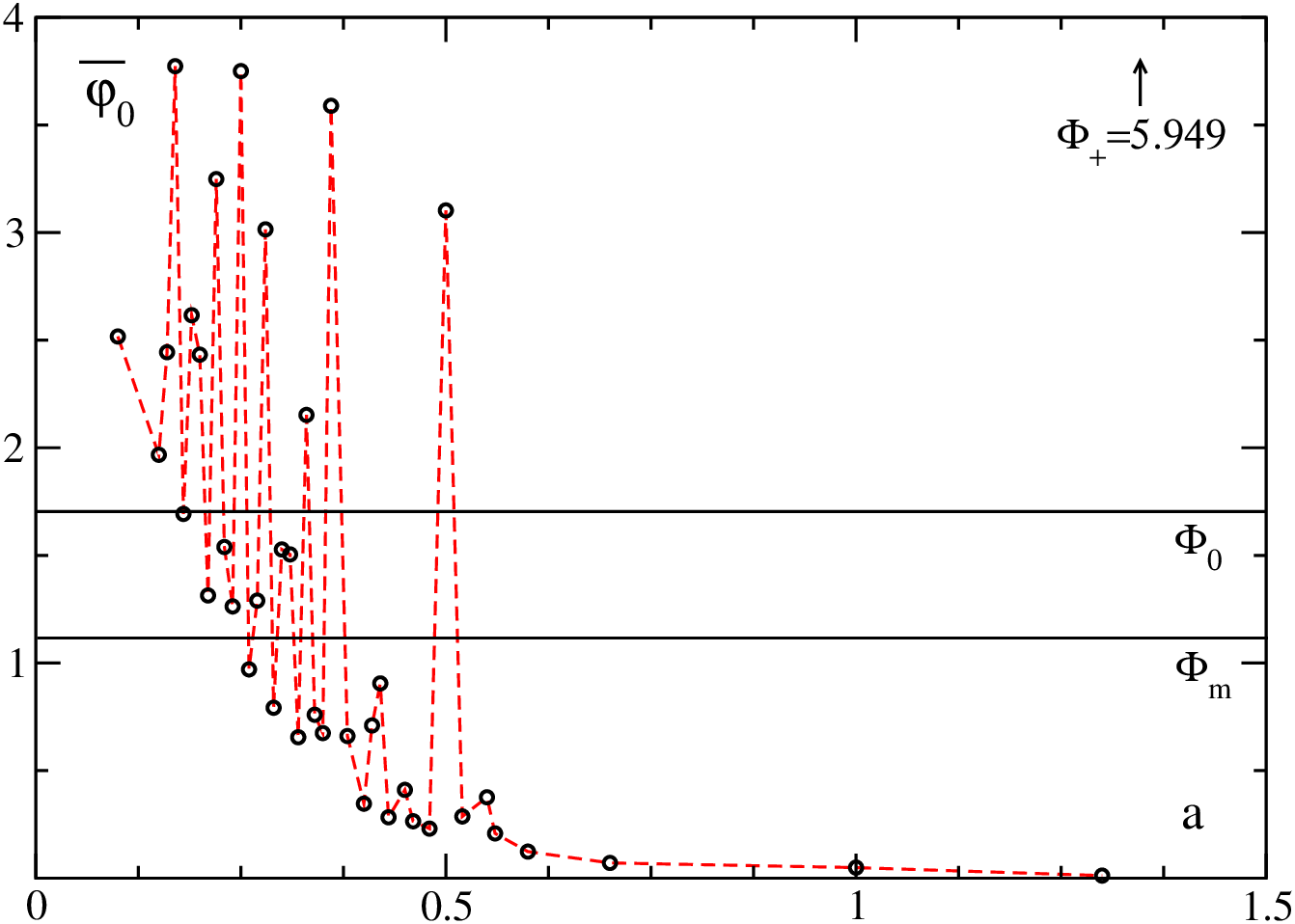}\vspace*{3mm}
\end{center}
\caption{\label{fig:0.6series}
The maximum $\bar \varphi_0$ of the expectation value
 $<\varphi_0(t)>$ for set $II$,
$\alpha=0.6,\beta=2$, as a function of $a$.}
\end{figure}

\begin{figure}
\begin{center}
\includegraphics[scale=0.4]{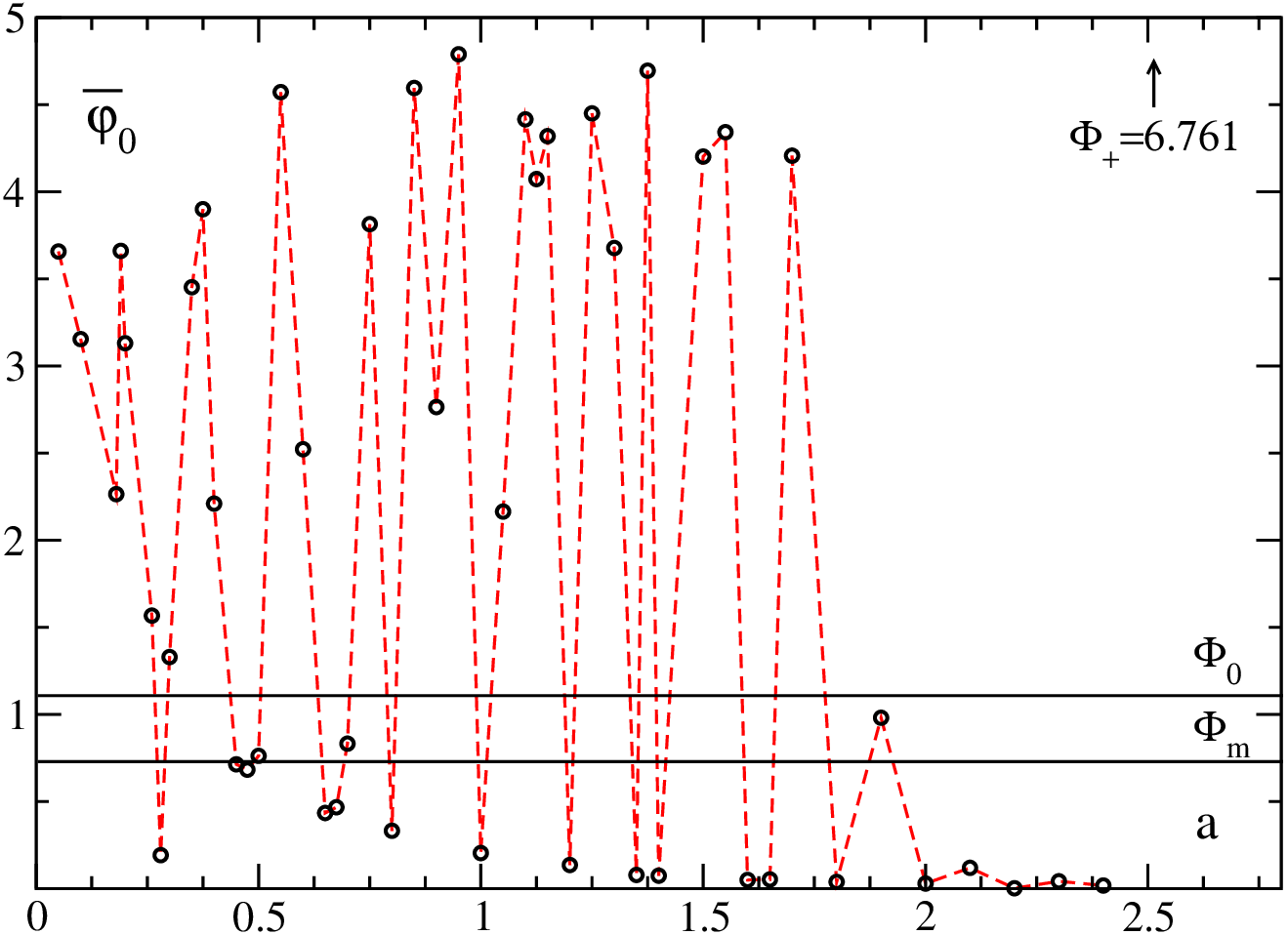}\vspace*{3mm}
\end{center}
\caption{\label{fig:0.4series}
The maximum $\bar \varphi_0$ of the expectation value
 $<\varphi_0(t)>$ for set $III$,$\alpha=0.4,\beta=1$,
as a function of $a$.}
\end{figure}

\begin{figure}
\begin{center}
\includegraphics[scale=0.4]{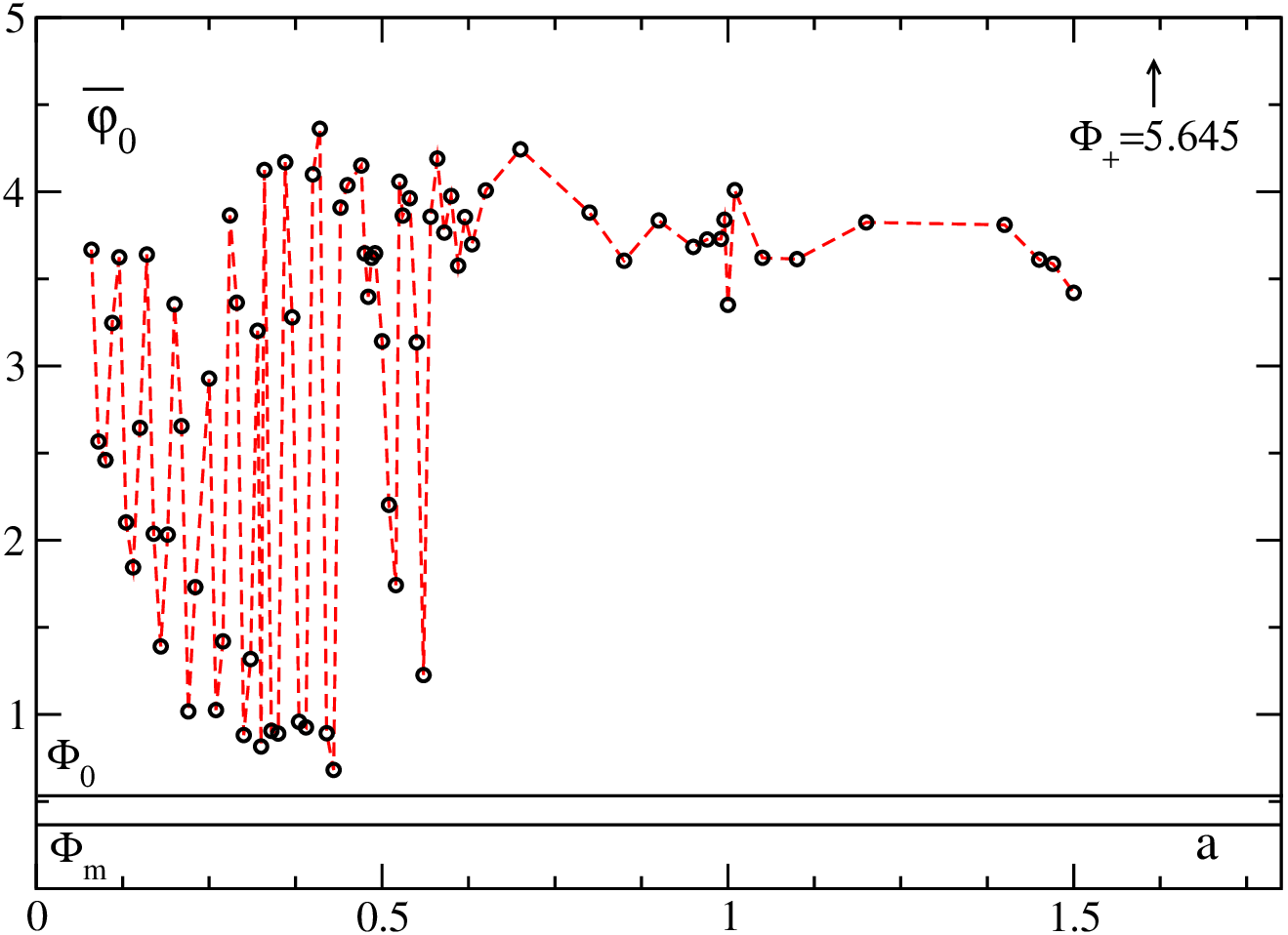}\vspace*{3mm}
\end{center}
\caption{\label{fig:0.25series}
The maximum $\bar \varphi_0$ of the expectation value
 $<\varphi_0(t)>$ for set $IV$,$\alpha=0.25,\beta=0.25$,
as a function of $a$.}
\end{figure}

For the parameter sets $I-III$ the tunneling shuts off entirely at
higher values of $a$ as expected from the bounce
action \eqn{homogeneousbounceaction} being proportional to
$a$. As large $\beta$ implies a large classical
action and, therefore, a small semiclassical tunneling rate,
this shutting-off already happens at small values,  $a \simeq 0.5$, for set
$II$. For set $I$ on still finds resonant tunneling for $a > 3$. However,
the transition time at the resonances increases substantially
($ t \simeq 1000$ at the last  peak in Fig. 1), between 
the resonances the amplitude of the
zero mode oscillations decays exponentially with $a$, as for 
sets $II$ and $III$, and
the zero mode wave function displays no tunneling at all.
 The case of set $IV$ is quite different. Here at large
$a$ the system always tunnels. This effect is due to the fact
that the quantum fluctuations deform the potential seen by 
the zero mode, Eq. \eqn{effectivezeromodepotential} 
into a potential without barrier, the late time
wave function extending over a range from $-5$ to $+20$, we call
this kind of transition a sliding transition. We will discuss this
below.

The results for the maximal value $\bar\varphi_0$ of the zero mode average
display a global picture of the tunneling
in  the asymmetric double well potential in quantum
field theory. They look quite different from what one
usually assumes to happen in ``false vacuum decay'', when one just
considers the  homogeneous bounce action \eqn{homogeneousbounceaction}
without or with quantum corrections.
We will now discuss in some more detail the evolution of the
zero mode wave function. 

For all parameter sets and all values
of $a$ the wave function of the zero mode initially evolves slowly;
it becomes slightly asymmetric and develops a tail into the region
of the potential barrier. Here already the fact that we do not restrict it
to be Gaussian is of prime importance. The further behavior then depends
on the parameter sets.

\begin{figure}
\begin{center}
\includegraphics[scale=0.4]{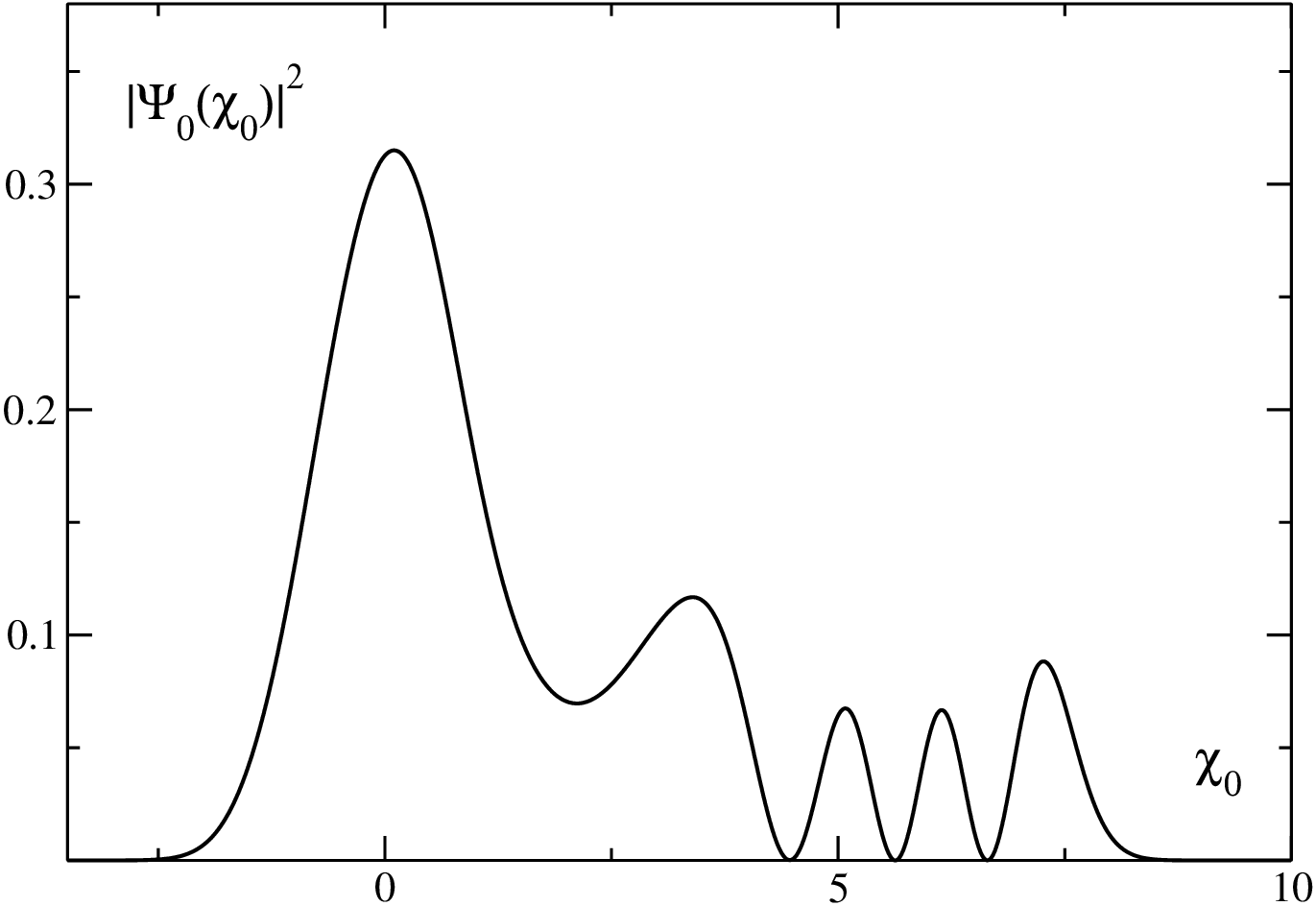}\vspace*{3mm}
\end{center}
\caption{\label{fig:wf.8.51.35} Probability density $|\psi_0(\chi_0)|^2$
at intermediate times, here $t=10$, for $\alpha=0.8$, $\beta=0.5$, and
$a=1.35$}
\end{figure}

{\em On the resonances}
 the wave function, which initially is essentially
the ground state wave function of the left well first connects to a
particular wave function in the right well, obviously the
one of the degenerate level. This is seen in Fig. \ref{fig:wf.8.51.35}   , the
well-defined number of peaks within the right hand well
indicate a specific approximate level within the right hand
well. If the fluctuations of the quantum field,
i.e., the amplitudes of the $\varphi_n$, remain small, then
one observes an oscillation forth and back between these
two wave functions, as expected in a purely quantum mechanical
system. If the nonzero field modes get excited, then these
oscillations of the zero mode are disturbed, other wave functions mix in,
as seen by the increasing number of peaks, in some cases the wave
function looks quite chaotic.

{\em Off resonance}
 the expectation value of $\varphi_0$ oscillates regularly
with a small amplitude. It still develops a tail into the right hand well,
but this part of the wave functions remains small.

{\em Beyond the resonance region} there is efficient tunneling
if the fluctuations of the nonzero momentum modes are large.
In this case, while the wave function penetrates into the 
barrier region, the fluctuation integral becomes positive, the
barrier disappears and the wave function slides down the hill
retaining almost its initial Gaussian form.

\begin{figure}
\begin{center}
\includegraphics[scale=0.4]{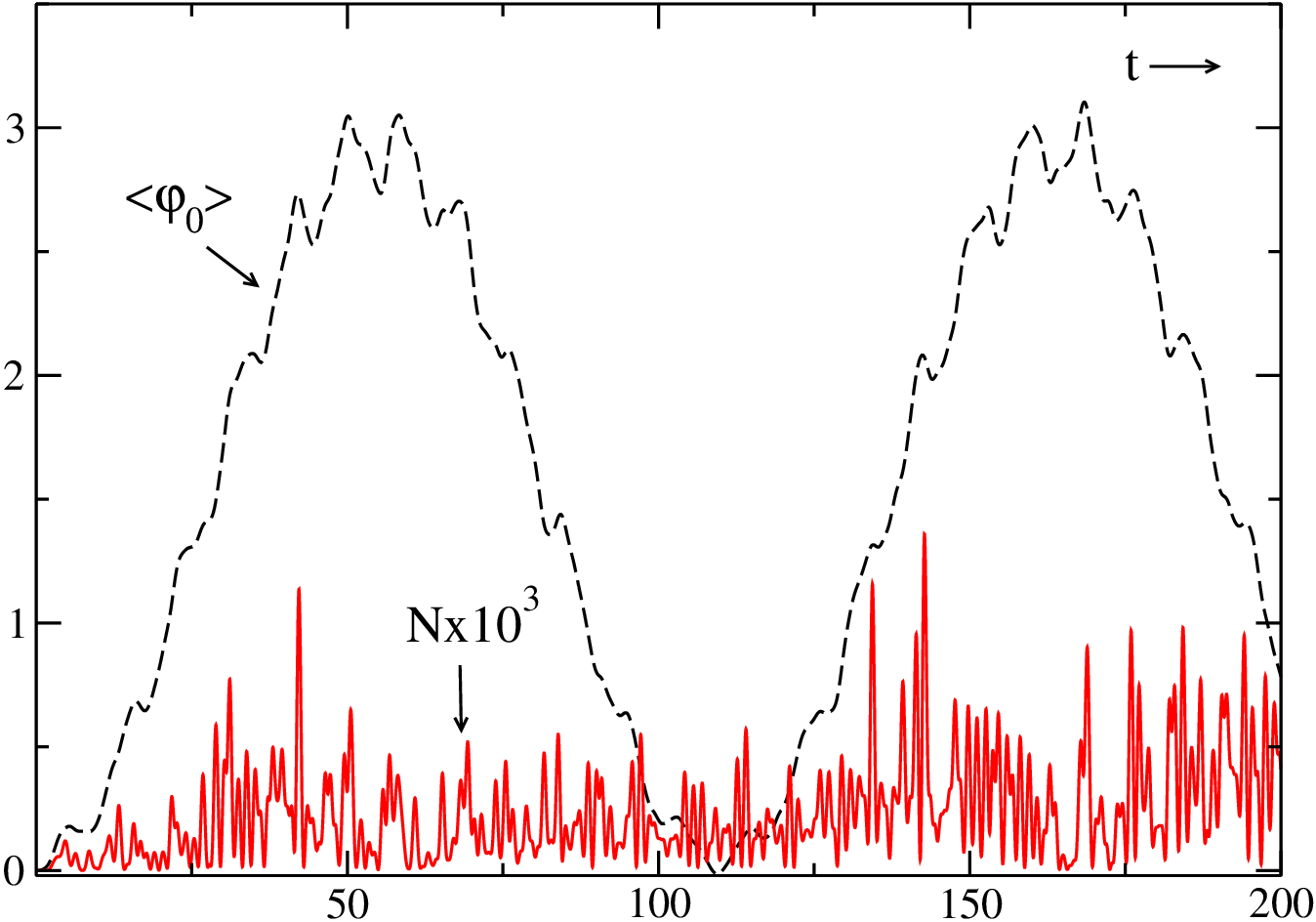}\vspace*{3mm}
\end{center}
\caption{\label{fig:zm0.620.5}
Evolution of the expectation value of the zero mode
$<\varphi_0>$ and of the particle number $\cal N$; parameters
$\alpha=0.6$, $\beta=2$, $a=0.5$.}   
\end{figure}

\begin{figure}
\vspace*{6mm}\begin{center}
\includegraphics[scale=0.4]{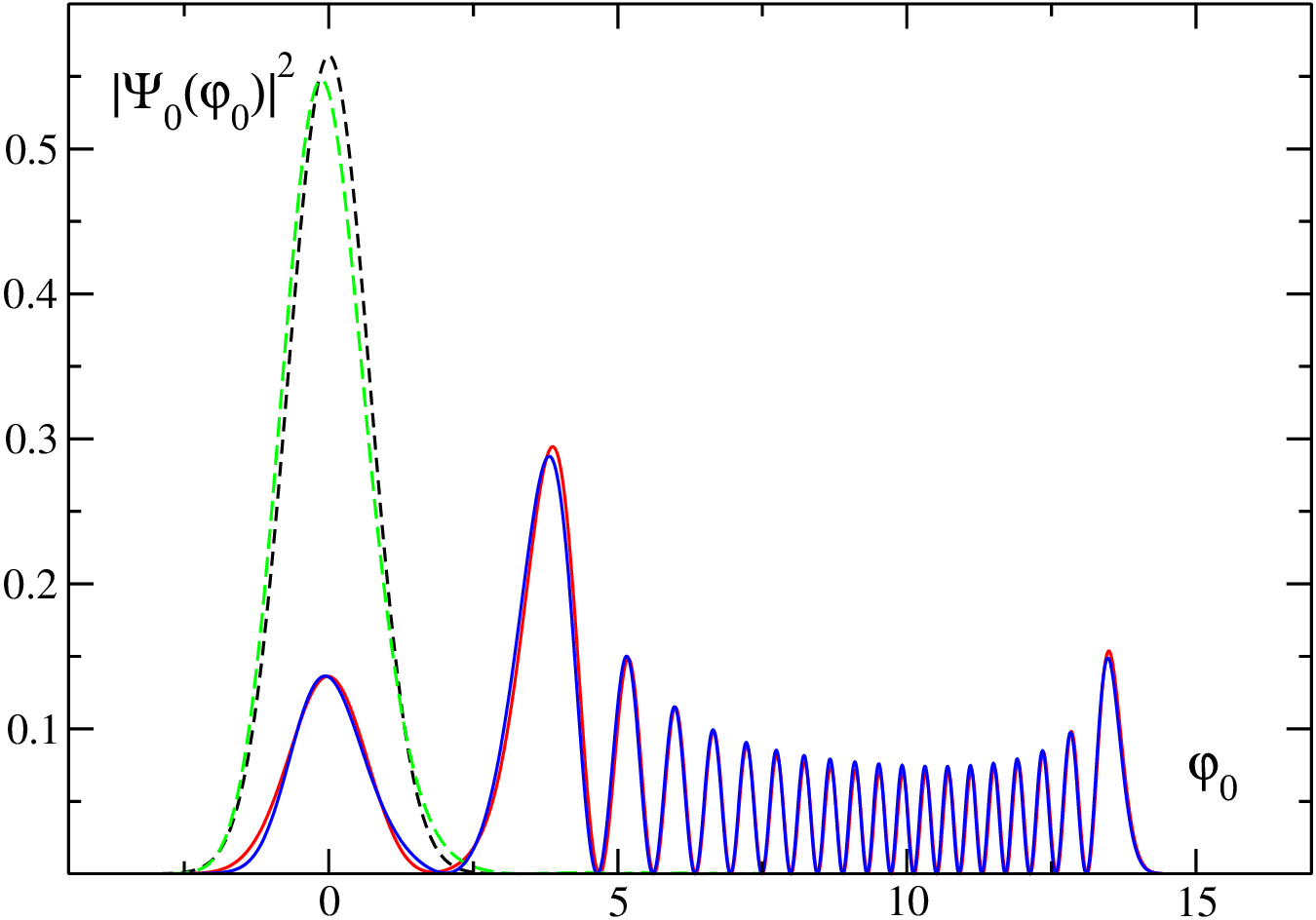}\vspace*{3mm}
\end{center}
\caption{\label{fig:wf0.620.5}
The zero mode wave function for $\alpha=0.6$, $\beta=2$, $a=0.5$;
dashed lines: the wave functions at $t=0$ and $109$; solid lines:
the wave functions at $t=50$ and $t=168$.}
\end{figure}

The simulations of set II, $\alpha=0.6, \beta=2$ are expected to
resemble most closely the case of quantum mechanics. The nonzero
momentum quantum 
fluctations indeed remain small and we have the resonances expected
from the qualitative picture of degenerate levels of the individual 
wells. The results for the expectation value of the zero mode
and for the evolution of the wave functions correspond to these expectations.
We present the regular oscillations of the zero mode on the
resonance at $a=0.5$ in Fig. \ref{fig:zm0.620.5}. We plot, in the same figure,
the particle number $\caln$ defined below Eq. \eqn{particlenumbers},
multiplied by $1000$. The wave function, shown at four different times
in Fig. \ref{fig:wf0.620.5}, is seen to return to itself almost exactly
after half a period of $t\simeq 55$.

Set III is intermediate in the sense that with $\beta=1$ we expect
the fluctuations of the nonzero modes to be sizeable but not really large.
We find that tunneling is again characterized by the occurrence of
resonances, and again shuts off at larger values of $a$, here around 
$a=2$. Around this value the resonant transitions occur on longer
and longer time scales. So for $a>2$ we cannot exclude further
resonances if we run the simulations for times longer than
$t\simeq 300$.   

\begin{figure}
\begin{center}
\includegraphics[scale=0.4]{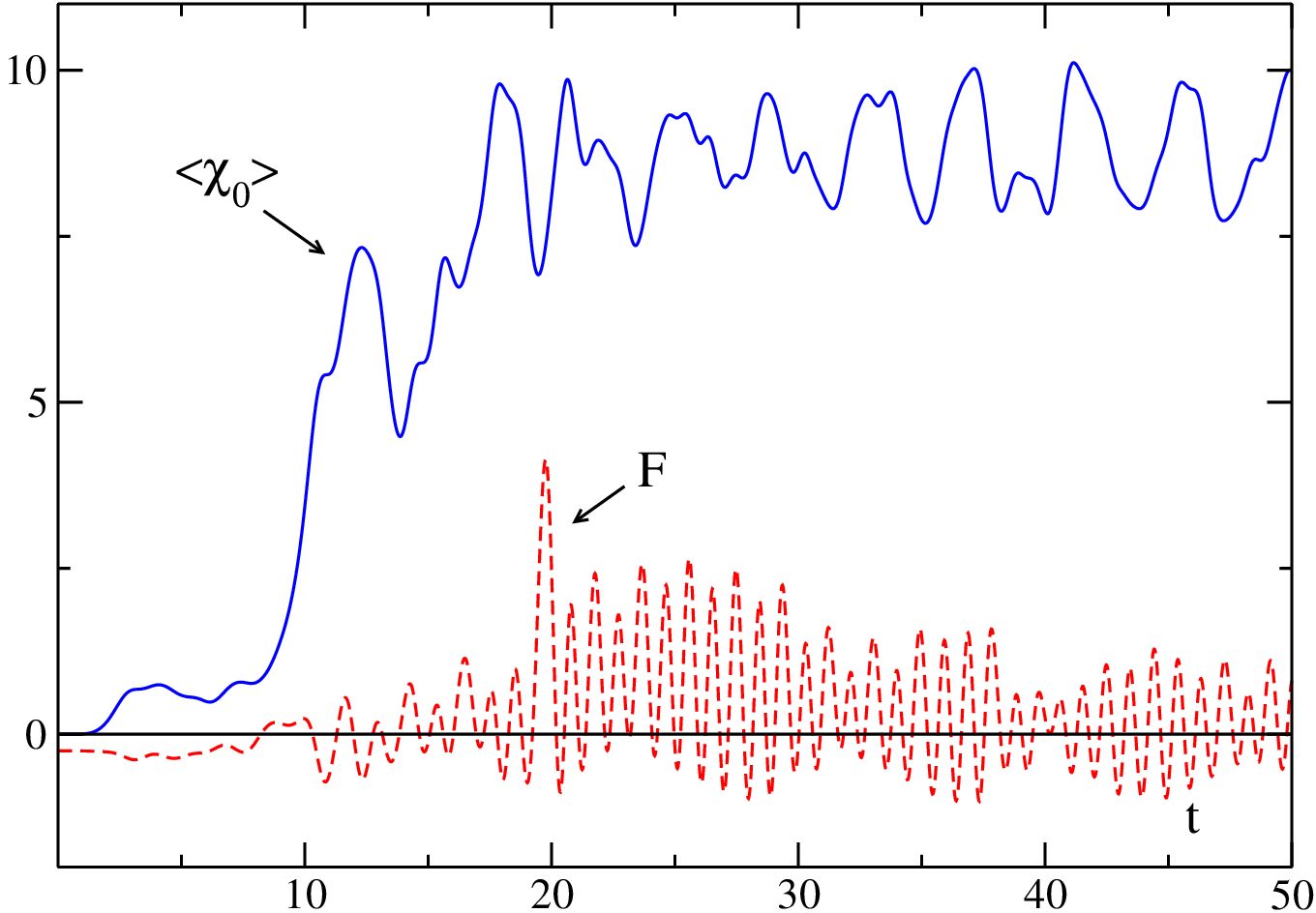}
\end{center}
\caption{\label{fig:zm1111.2}
Evolution of the expectation value $<\varphi_0(t)>$ and of the
fluctuation integral $\calf(t)$ for $\alpha=\beta=0.25$ (set IV)
and $a=1.2$.}
\end{figure}

\begin{figure}

\begin{center}
\includegraphics[scale=0.4]{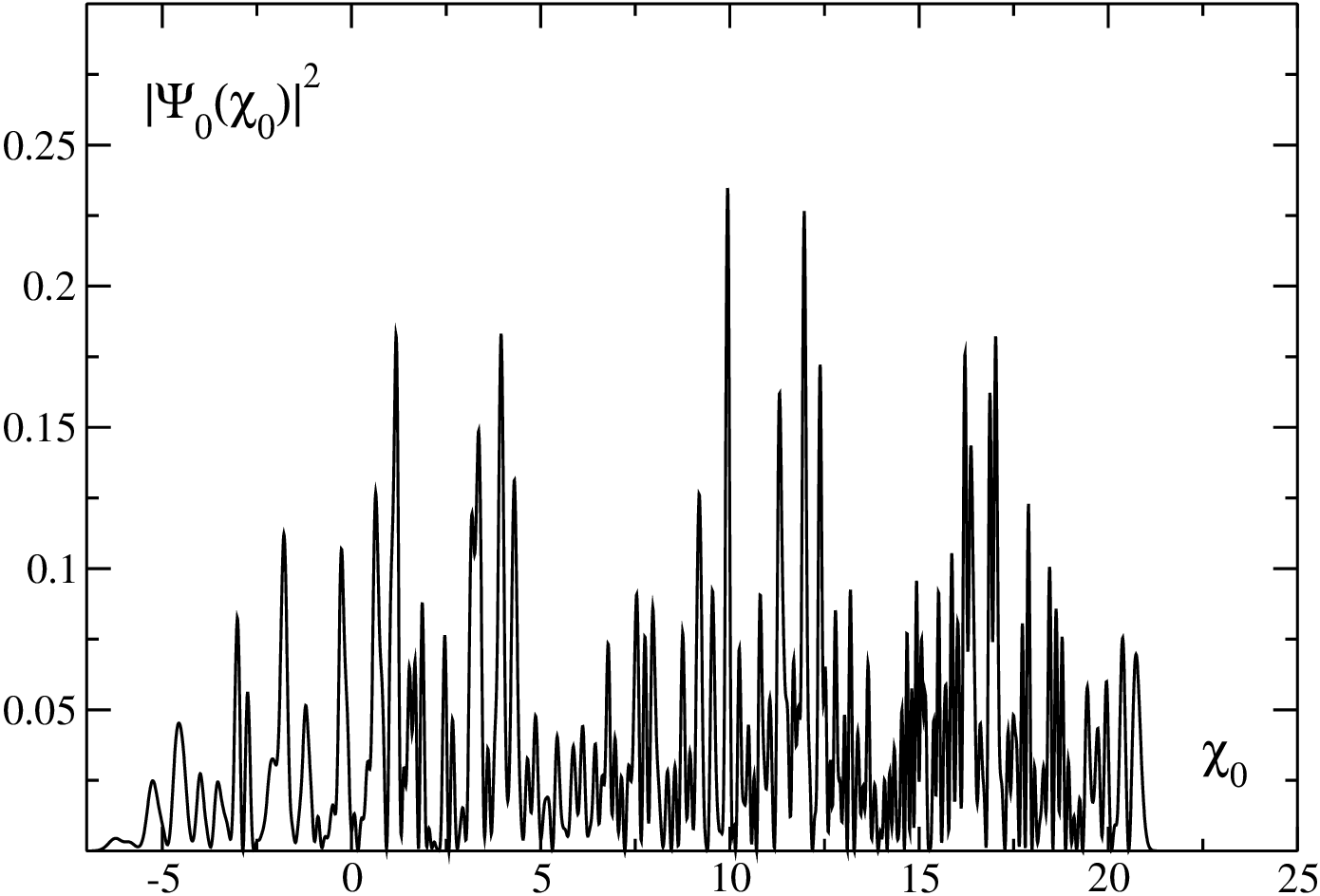}
\end{center}
\caption{\label{fig:wf1111.2}
The wave function after tunneling, $\alpha=\beta=0.25$ (set IV)
and $a=1.2$ and time $t=30$.}
\end{figure}

For set IV the zero mode shows resonant behaviour at small 
$a$, but for $a > 0.5$ the system always ends up in the
right well. The actual evolution is quite involve here.
Once the fluctuations set in, they tilt the potential in such
a way that the barrier disappears and 
that the wave function can start to move right. 
Then the fluctuation
integral $\calf$ gets small again or negative, so the potential 
barrier appears again. This process repeats itself in an
oscillatory way, and the wave function gradually shifts towards
the deeper well. This is displayed in Fig. \ref{fig:zm1111.2}.
At late times the quantum modes still oscillate with a
sizeable amplitude, so that
the potential seen by the zero mode keeps on changing
and can neither be considered to be double or single well.
A typical wave function at late times is shown in Fig. \ref{fig:wf1111.2}
for $a=1.2$ and $t=30$. One sees no trace of a double-well structure.
In spite of the complexity of the wave function the energy is still
conserved better than one part in $10^8$ (see Fig. \ref{fig:energy}). 


\begin{figure}
\begin{center}
\includegraphics[scale=0.4]{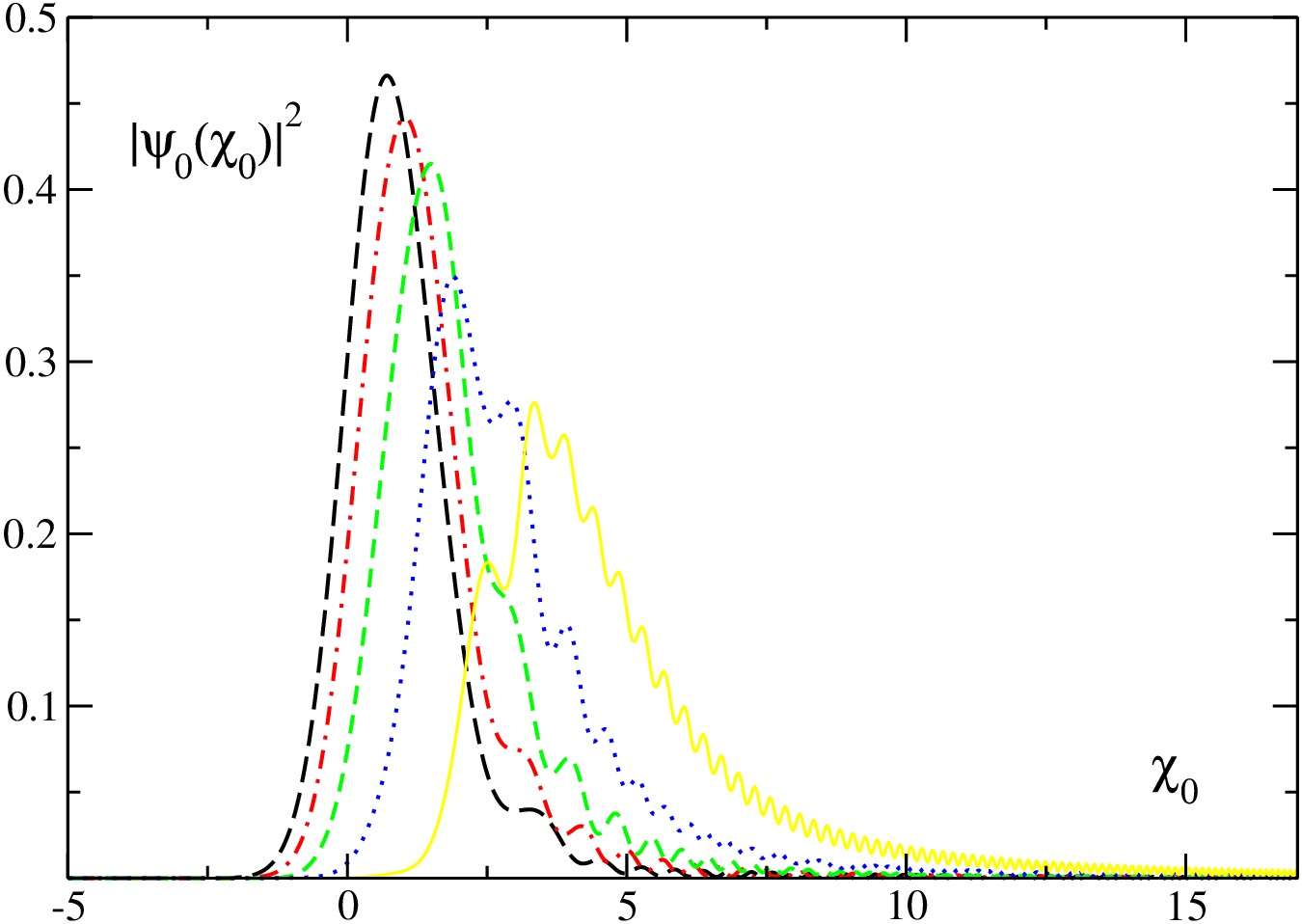}
\end{center}
\caption{\label{fig:psi_slidea6}
Evolution of the wave function during the barrier transition:
parameters $m=\eta=\lambda=1$; $a=6$; long-dashed line: $t=7$;
dash-dotted line: $t=7.4$; short-dashed line: $t=7.8$; dotted line:
$t=8.2$; straight line: $t=8.6$.}
\end{figure}

The situation becomes more transparent for larger values
of $a$ \footnote{These values do not appear in Fig.
\ref{fig:0.25series} as otherwise the structure in the resonance region
cannot not be resolved.}.
 We show results for $a=6$. Here again the potential
is tilted by the fluctuations, but the fluctuation integral
remains positive on the average at late times. This implies that
the barrier has disappeared definitively. The transition 
can be described as a {\em sliding} of the wave function
towards the new minimum. This is displayed in Fig. \ref{fig:psi_slidea6}. 

\begin{figure}
\begin{center}
\includegraphics[scale=0.4]{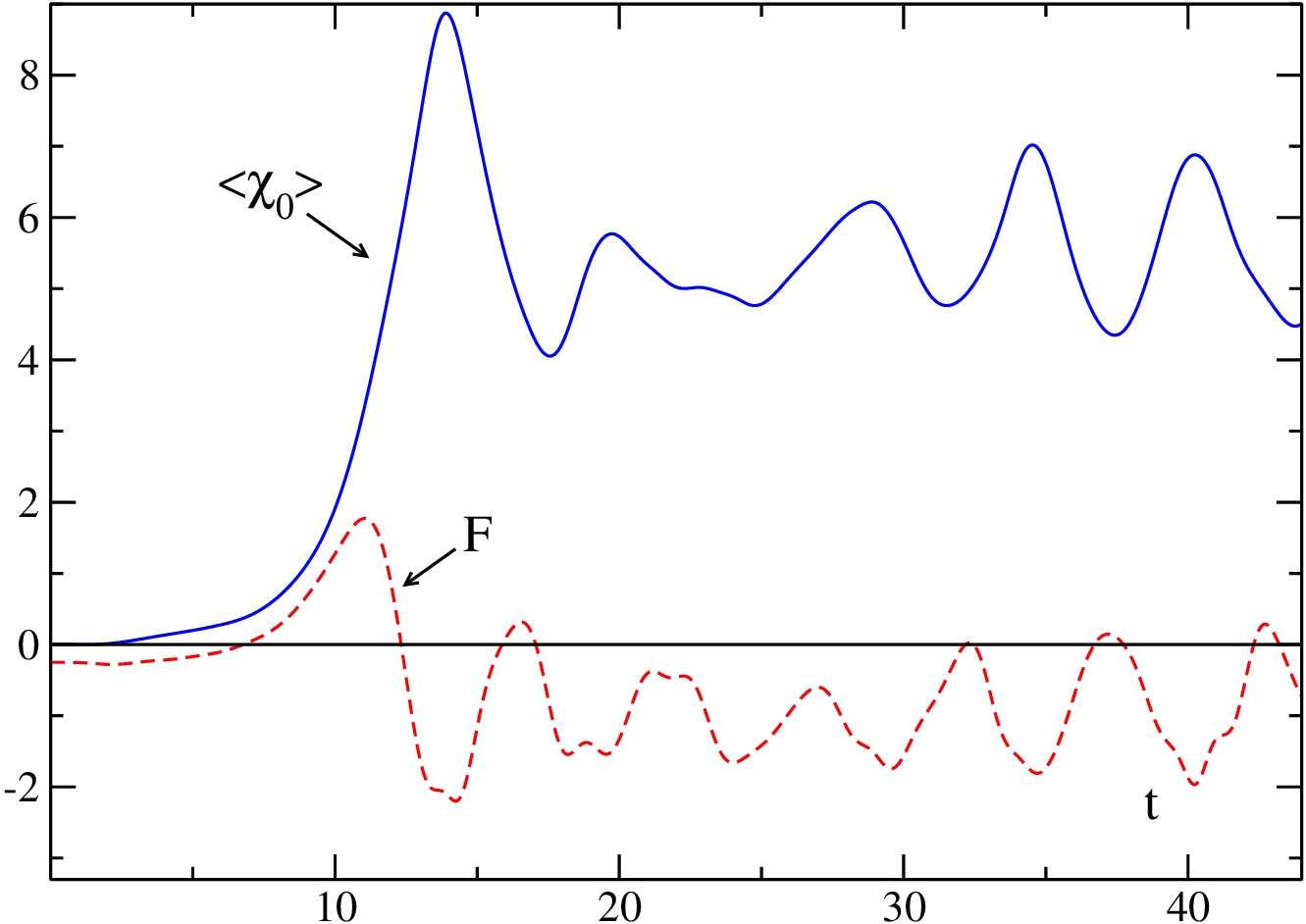}
\vspace*{3mm}
\end{center}
\caption{\label{fig:zm.8.25}
Evolution of the expectation value $<\varphi_0(t)>$ and of the fluctuation
integral for $\alpha=0.8$, $\beta=0.2$, and $a=5$.}
\end{figure}

\begin{figure}
\begin{center}
\vspace*{3mm}
\includegraphics[scale=0.4]{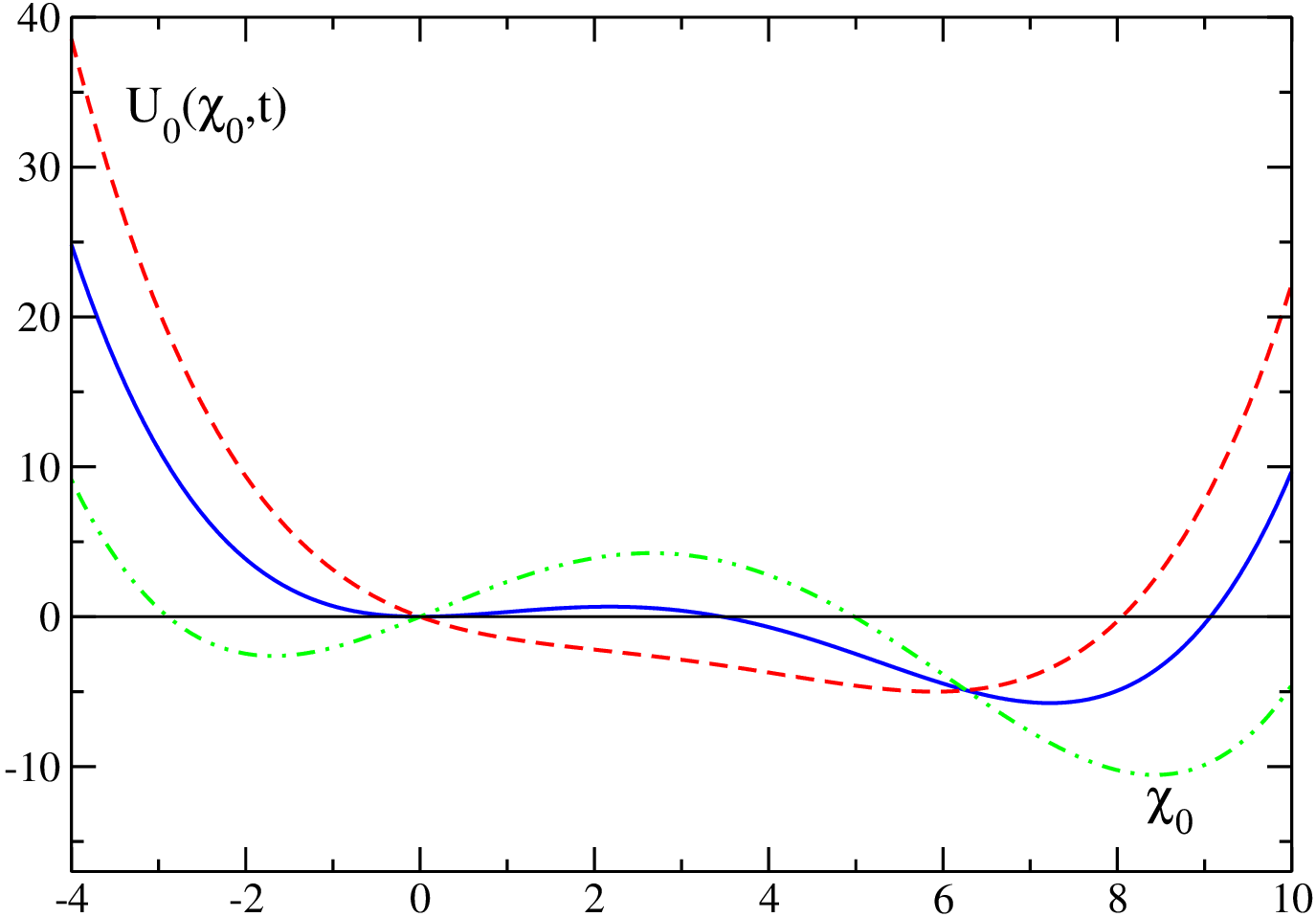}
\vspace*{3mm}
\end{center}
\caption{\label{fig:pot.8.25}
Evolution of the effective potential $U_0(\chi_0,t)$;
solid line: $t=0$; dashed line: $t=11$; dash-doubledotted line:
$t=13$.}
\end{figure}

For the case under consideration, set $IV$ with $\alpha=0.25 $ 
the potential is very asymmetric. So one may
infer that the tilting of the potential is most effective here,
and that this effect may in fact be limited to small values of $\alpha$.
We therefore have performed simulations for $\alpha=0.8$ and 
$\beta=0.2$, i.e., for a more symmetric potential, but again with
large quantum corrections. We find that for values of $a > 1$  
the transition again occurs by sliding, but that at late times the
fluctuation integral becomes negative and the potential is tilted
in such a way that the barrier reappears. This is presented in Figs.
\ref{fig:zm.8.25} and \ref{fig:pot.8.25}. The wave function 
at late times is then found to have sizeable amplitudes in {\em both} wells.

So it seems that {\em sliding} occurs universally for small $\beta$
and large $a$.


\section{Conclusions and outlook}
\setcounter{equation}{0}
\label{outlook}

We have presented here an analysis of global tunneling
in quantum field theory on a compact space. 
This analysis was based on a real-time
formulation, using the time-dependent Hartree approximation.
We have performed numerical simulations of a system where
the wave function of the zero momentum mode evolves as a 
solution of the Schr\"odinger
equation, while the nonzero momentum modes are treated in the Gaussian
approximation. The approximation includes the back-reaction of the
nonzero momentum modes onto the zero momentum mode and
onto themselves. The renormalized equations were obtained
using the 2PPI formalism, adopted here to a system with finite
space extension. 

We have found that tunneling for such  systems occurs in a variety of 
different ways. For large $\beta$, implying small
field fluctations, the system behaves like a quantum
mechanical system with a single degree of freedom. Tunneling is effective
whenever it connects degenerate modes, so varying the parameters,
here the radius $a$, one finds resonance enhancements. For large 
spatial extension $a$
this resonant tunneling shuts off as expected from the fact
that the WKB integral over the barrier is proportional to $a$. For smaller
values of $\beta$ the nonzero momentum modes, the modes which make up
the quantum field, are enhanced and modify the quantum mechanical behavior.
If this enhancement is weak it can be considered as a kind of dissipation;
the mean value of the zero mode shows regular periodic oscillations, 
on and off resonance. On resonance this again involves two or a few more
modes in the two separate wells, off resonance the wave function 
essentially remains in the left well, with some tail in the right hand well.
For parameter sets where the excitation of the nonzero momentum
modes becomes sizeable, these regular oscillations 
are disturbed, the oscillations of
the mean value of the zero mode become irregular
and so do the wave functions of the zero mode. Off resonance
at late times the system again remains concentrated in the left hand well.
On resonance at late times the wave function
extends over the entire region allowed at the given energy. While the 
system may then be considered as having tunneled, it certainly ends up
in a rather complex excited state.

If the quantum corrections are large the system exhibits another phenomenon:
the quantum fluctuations tilt the effective potential for the
zero mode in such a way that the barrier disappears entirely.
In such cases the system wave function can be
observed to slide into the new minimum. The disappearance ot the 
barrier may be a temporary effect. It   is due to the substantial quantum
excitations once the wave function enters the barrier, leading to negative
squared masses for the fluctations. At late times the potential
may become be a double-well potential again, or the barrier may disappear 
defintively.

This variety of phenomena observed here in a real-time analysis 
cannot be expected to be described by a simple transition rate
formula. A better approximation than the classical bounce action, 
Eq. \eqn{homogeneousbounceaction}, could be obtained by fitting 
together WKB patches within the allowed and forbidden regions or by applying
a time-dependent WKB approximation to the $k=0$ mode itself, as done
for quantum mechanics in Ref. \cite{Cooper:1986wv}. However, the advantage
of dealing with analytic or semi-analytic formulas is lost at the latest
if one includes the nonzero momentum fluctuations. In most cases they cannot
be evaluated analytically and, as implied by the present analysis,
they cannot be expected to give reliable estimates. For homogeneous
tunneling they exhibit the phenomenon of multiple unstable modes: if the
unstable mode for $k=0$ has the squared frequency $-|\omega_-|^2 < 0$
then the $k\neq 0$  modes will have an eigenmode with the eigenvalue
$k^2/a^2-|\omega_-|^2$ which is positive for small $a$ but will become negative
whenever $a > a_k=k/|\omega_-|$. The interpretation of these modes
is unclear, there is no trace of them in the numerical simulations.

With increasing $a$ one expects the system to become unstable
with respect to inhomogeneous, local bounces. This instability may be
closely related to the sliding phenomenon for which a strong
excitation of $k\neq 0$ modes is crucial: the $k\neq 0$ modes
are of course spatially inhomogeneous; if they are excited substantially
they may be interpreted as {\em classical} fluctuations.
For $k=n$ they have the general form of a chain of $n$ bubbles,
much in the way as they are depicted in Refs. \cite{Linde:1980tt,Linde:1981zj}
for the decay of the false vacuum at {\em finite temperature}.
It could be interesting to investigate this relation between 
inhomogeneous and homogeneous tunneling in a more quantitative way.
Clearly, in view of this instability with respect to local bounces,
   our present analysis  becomes unreliable
at values  $a >> 1$.

The restriction to one compact space dimension
was done in order to reduce technical complications to a minimum.
Similar phenomena are expected for compact three-dimensional spaces. The 
discrete spectrum of momenta then becomes a discrete spectrum
of angular  momenta, whose excitation will be small for
small $a$ and become more effective with increasing $a$. The potential
seen by the homogeneous mode is again a double-well potential, so
again resonances are expected to dominate the low $a$ regime.
If gravity is included, like for transitions in de Sitter
space, renormalization becomes problematic, especially if quantum
backreaction is included. Still it would be interesting
to investigate along these lines the relation between homogeneous
transitions like the Hawking-Moss instanton and inhomogeneous
ones like the Coleman-deLuccia bounce.

Of course it would be even more interesting to follow, in real time,
the {\em local} creation of real vacuum bubbles
including the associated excitation of quantum modes with back-reaction.
This seems to be out of scope with the presently available
numerical methods and computer facilities. 
Our calculations may elucidate in a qualitative way
effects that are missed in the WKB approach to local transitions, even 
when quantum corrections are taken into
account \cite{Baacke:1993ne,Baacke:2003uw,Bergner:2003id,Baacke:2004xk}.

\section*{Acknowledgments}
One of us (N.K) thanks the Deutsche Forschungsgemeinschaft
for support in the framework of  the Graduiertenkolleg 841:
``The physics of elementary particles at accelerators and in the universe''.

\bibliography{qtpap}
\bibliographystyle{h-physrev4}

\end{document}